\documentclass[a4paper,11pt]{article}
\usepackage{jheppub} 
\usepackage{lineno}

\usepackage{amsmath,amssymb}
\usepackage{graphicx}
\usepackage[normalem]{ulem}
\usepackage[dvipsnames]{xcolor}
\usepackage{cancel}
\usepackage[T1]{fontenc}
\usepackage{url}
\usepackage{balance}
\usepackage{xspace}
\usepackage{topcapt}

\newcommand{\gve}{\ensuremath{g_{ve}}\xspace}
\newcommand{\MX}{\ensuremath{M_X}\xspace}
\newcommand{\pot}{POT\xspace}
\newcommand{\npot}{\ensuremath{N_{\mathrm{POT}}}\xspace}
\newcommand{\longnpot}{\ensuremath{N_{\mathrm{POT}}(s)}\xspace}
\newcommand{\Ntwo}{\ensuremath{N_2}\xspace}
\newcommand{\longNtwo}{\ensuremath{N_2(s)}\xspace}
\newcommand{\com}{CoM\xspace}
\newcommand{\cog}{CoG\xspace}
\newcommand{\sqrts}{\ensuremath{\sqrt{s}}\xspace}
\renewcommand{\S}{\ensuremath{S}\xspace}
\newcommand{\longS}{\ensuremath{S(s; \MX, \gve)}\xspace}
\newcommand{\eps}{\ensuremath{\varepsilon_{\text{sig}}}\xspace}
\newcommand{\longeps}{\ensuremath{\varepsilon_{\text{sig}}(s)}\xspace}
\newcommand{\B}{\ensuremath{B}\xspace}
\newcommand{\longB}{\ensuremath{B(s)}\xspace}
\newcommand{\gR}{\ensuremath{g_R}\xspace}
\newcommand{\longgR}{\ensuremath{g_R(s)}\xspace}
\newcommand{\K}{\ensuremath{K}\xspace}
\newcommand{\longK}{\ensuremath{K(s)}\xspace}
\newcommand{\Rmin}{\ensuremath{R_{\text{min}}}\xspace}
\newcommand{\Rmax}{\ensuremath{R_{\text{max}}}\xspace}

\newcommand{\eg}{\mbox{e.g.}\xspace}     
\newcommand{\vs}{\mbox{vs.}\xspace}      
\providecommand{\NA}{\ensuremath{\text{---}}}    

\newcommand{\BabaYaga}{\textsc{BabaYaga}\xspace}
\newcommand{\GEANTfour} {{\textsc{Geant4}}\xspace}

\newcommand{\abs}[1]{\ensuremath{\lvert #1 \rvert}}
\newcommand{\CL}{\ensuremath{\text{CL}}\xspace} 
\newcommand{\CLs}{\ensuremath{\text{CL}_\text{s}}\xspace}
\newcommand{\CLsb}{\ensuremath{\text{CL}_\text{s+b}}\xspace}

\newlength\TabSkip
\newlength\BigSkip

\newcommand{\unit}[1]{\ensuremath{\text{\,#1}}\xspace}
\newcommand{\MeV}{\ensuremath{\,\text{Me\hspace{-.08em}V}}\xspace}
\newcommand{\micron}{\ensuremath{\,\mu\text{m}}\xspace}
\newcommand{\mm}{\ensuremath{\,\text{mm}}\xspace}
\newcommand{\mus}{\ensuremath{\,\mu\text{s}}\xspace}

\arxivnumber{2505.24797} 

\title{\boldmath Search for a new 17~MeV resonance via $e^+e^-$ annihilation with the PADME Experiment}

\author[a]{{\bf The PADME Collaboration} \\ F. Bossi}
\author[a]{R. De Sangro}
\author[a]{C. Di Giulio}
\author[a]{E. Di Meco}
\author[a]{D. Domenici}
\author[a]{G. Finocchiaro}
\author[a]{L.G. Foggetta}
\author[a]{M. Garattini}
\author[a]{P. Gianotti}
\author[a]{M. Mancini}
\author[a]{I. Sarra}
\author[a]{T. Spadaro%
\footnote{Corresponding author.}}
\author[a]{C. Taruggi%
\footnote{At present working for Rome Technopole project}}
\author[a]{E. Vilucchi}
\author[b]{K. Dimitrova}
\author[b]{S. Ivanov}
\author[b]{Sv. Ivanov}
\author[b]{K. Kostova}
\author[b,a]{V. Kozhuharov}
\author[b]{R. Simeonov}
\author[c]{F. Ferrarotto}
\author[c]{E. Leonardi}
\author[c]{P. Valente}
\author[c,d]{E. Long}
\author[c,d]{G.C. Organtini}
\author[c,d]{M. Raggi}
\author[e]{A. Frankenthal}

\affiliation[a]{{INFN Laboratori Nazionali di Frascati, Via E. Fermi, 54 I-00044 Frascati, Italy}}
\affiliation[b]{{Faculty of Physics, Sofia University ``St. Kl. Ohridski'', 5 J. Bourchier Blvd., BG-1164 Sofia, Bulgaria}}
\affiliation[c]{{INFN Sezione di Roma, p.le Aldo Moro 5, I-00185 Rome, Italy}}

\affiliation[d]{{Physics Department, ``Sapienza'' Universita di Roma, p.le Aldo Moro 5, I-00185 Rome, Italy}}
\affiliation[e]{{Department of Physics and Astronomy, University of California, Irvine, Irvine, CA 92697-4575, USA}}

\abstract{
The PADME Experiment at the Frascati DA$\Phi$NE linear accelerator has searched for a hypothetical particle with mass around 17\MeV, commonly referred to as the X17, using a positron beam incident on a fixed  target. The beam energy was varied between 262 and 296\MeV, corresponding to center-of-mass energies \sqrts between 16.4 and 17.4\MeV. The X17 should be produced resonantly via $e^+e^-$ annihilation when \sqrts approaches its mass, inducing an excess of events with a two-body final state over the background expectation. The beam energy spacing was fixed to less than half the expected width of the resonance's line shape. Uncertainties below 1\% per \sqrts point were achieved. A blind analysis has been performed. The data are consistent with the expected background in most of the explored energy range, and limits are set in previously unexplored regions of the available parameter space. The most significant deviation is found for $\sqrts \approx 16.90\MeV$, corresponding to a global significance of approximately 2 standard deviations over the null hypothesis expectation.
}

\begin{document}
\maketitle

\section{Introduction}

In the last ten years, several experiments conducted with the ATOMKI pair spectrometer at Debrecen, Hungary, revealed anomalies in the distribution of opening angles between electron and positron pairs produced by the decay of excited states of $^{8}$Be, $^{12}$C and $^{4}$He nuclei~\cite{Krasznahorkay:2015iga, Krasznahorkay:2021joi, Krasznahorkay:2022pxs}. An analogous experiment conducted at VINATOM, in Vietnam, yielded similar results~\cite{Anh:2024req}. These anomalies could be explained by the existence of a new particle with mass around 17\MeV, the ``X17'', having a coupling \gve to electrons and positrons with strength of roughly $10^{-4}$ to $10^{-3}$~\cite{Feng:2016jff}, which corresponds to the following Lagrangian density~\cite{Darme:2022zfw}: 

\begin{equation}
    {\cal L}\supset \gve \, X^\mu_{17} \, \overline{e} \, \gamma_\mu \, e.
\end{equation}

More recently, the MEG II Collaboration at PSI attempted to replicate the original Beryllium result from ATOMKI, but did not observe a significant signal~\cite{MEGII:2024urz}. However, this result is still compatible with the ATOMKI observation at the level of 1.5 standard deviations ($\sigma$)~\cite{Barducci:2025hpg}.

Given its coupling to charged leptons, the X17 could be produced resonantly in electron-positron collisions at a center-of-mass (\com) energy equal to the X17 mass. Therefore, its existence could be inferred from an excess in the $e^+e^-\to e^+e^-$ production rate as a function of \com energy, when compared to the Standard Model (SM) prediction~\cite{Darme:2022zfw}. 

The PADME Collaboration at INFN's National Laboratories of Frascati has measured the production rate of the $e^+e^-\to e^+e^-$ and $e^+e^-\to\gamma\gamma$ processes for \com energies in the range $16.4<\sqrts<17.4\MeV$ during a three-month data-taking period in late 2022, covering the relevant mass region indicated by the ATOMKI results. We refer to this as the Run III dataset.  

At the relatively low energies of interest for the experiment, the motion of target atomic electrons considerably broadens the \com energy of the collisions and consequently the expected line shape of the X17 signal~\cite{Arias-Aragon:2024qji}. A specific procedure was designed and implemented to preserve the principles of a blinded data analysis while still enabling the validation of background predictions~\cite{Bertelli:2025mil}.

This paper reports the details of the experimental setup, data analysis strategy, experimental corrections and uncertainties, and observed results.
\section{The PADME setup}

PADME uses a beam of positrons incident on a fixed target and operates at CoM energies of $14 < \sqrts < 23 \MeV$~\cite{Raggi:2014zpa, Raggi:2015gza}. The experimental setup used in the Run III data-taking campaign is discussed below.

\subsection{Beam line}

A primary beam of positrons with a repetition rate of 49\unit{Hz} and custom energy selection is provided by the DA$\Phi$NE LINAC and Beam Test Facility 1 (BTF-1) complexes~\cite{Valente:2016tom,Valente:2017mnr,Foggetta:2021gdg}. A sketch of the beam line configuration is shown in figure~\ref{fig:accelerators}. For the Run III data taking, several machine parameters were optimized. A detailed discussion is reported in ref.~\cite{Bertelli:2024ezd}. Here we summarize the main beam adjustments:

\begin{figure}[ht]
\centering
\includegraphics[width=0.8\textwidth]{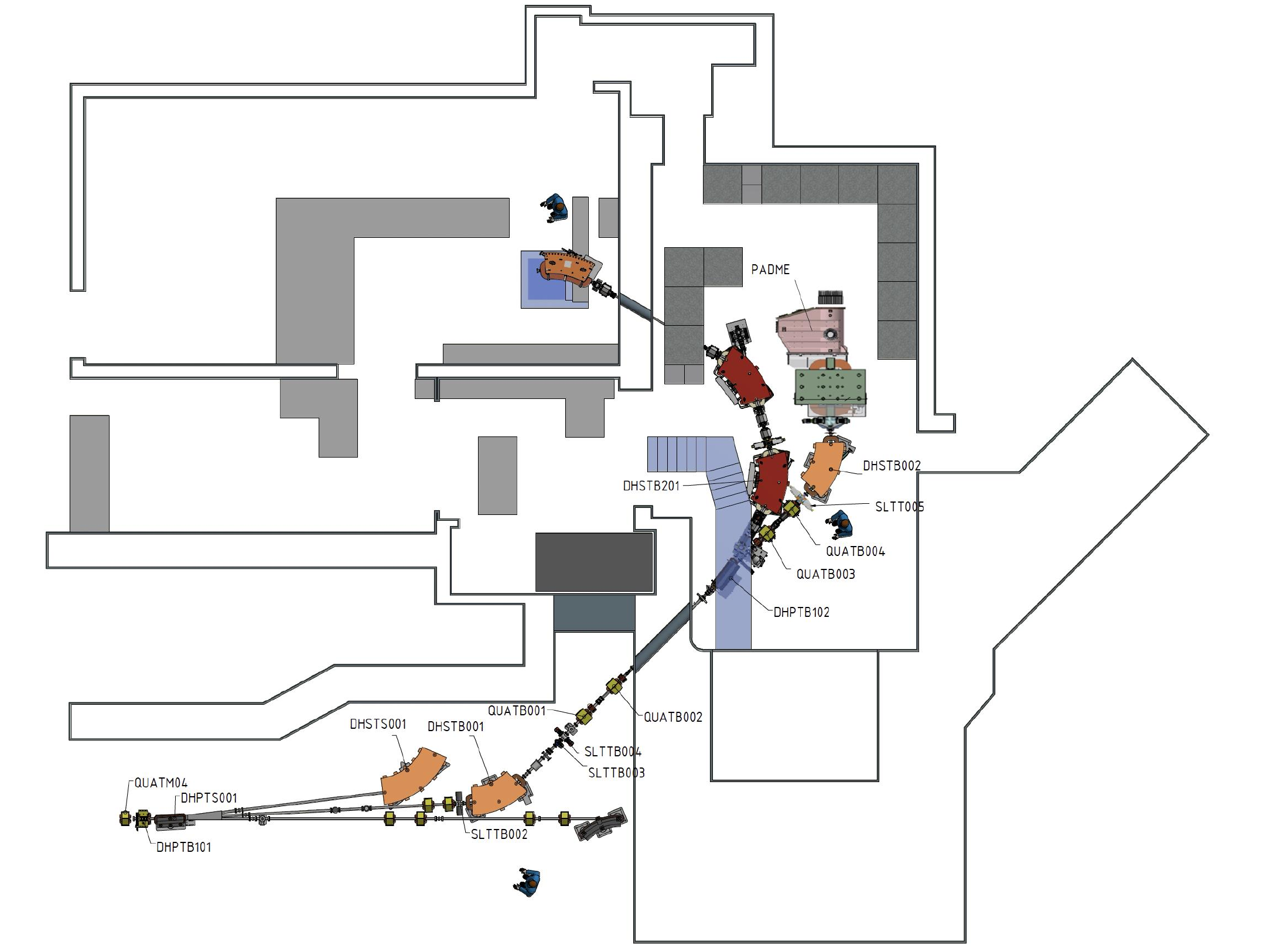}
    \caption{Sketch of the PADME beam line and the BTF-1 complex at LNF.} 
    \label{fig:accelerators}
\end{figure}

\begin{itemize}
\item Energy varied in the range 262--296\MeV (resonance scan), in addition to off-resonance points in the range 205--211\MeV and at 402\MeV;
\item Intensity lowered to roughly 3,000 positrons per bunch;
\item Spill duration set to approximately 200\unit{ns}, resulting in a relative beam energy spread at the detector of 0.25\%~\cite{Bertelli:2024ezd} because of the energy selection provided by the BTF beam line (bending magnets and collimators);
\item Emittance set to around $1\unit{mm}\times 1\unit{mrad}$ in both transverse directions~\cite{Buonomo:2023pzi}.

\end{itemize}

\subsection{Detector}

\begin{figure}[t]
\centering
\includegraphics[width=0.7\textwidth]{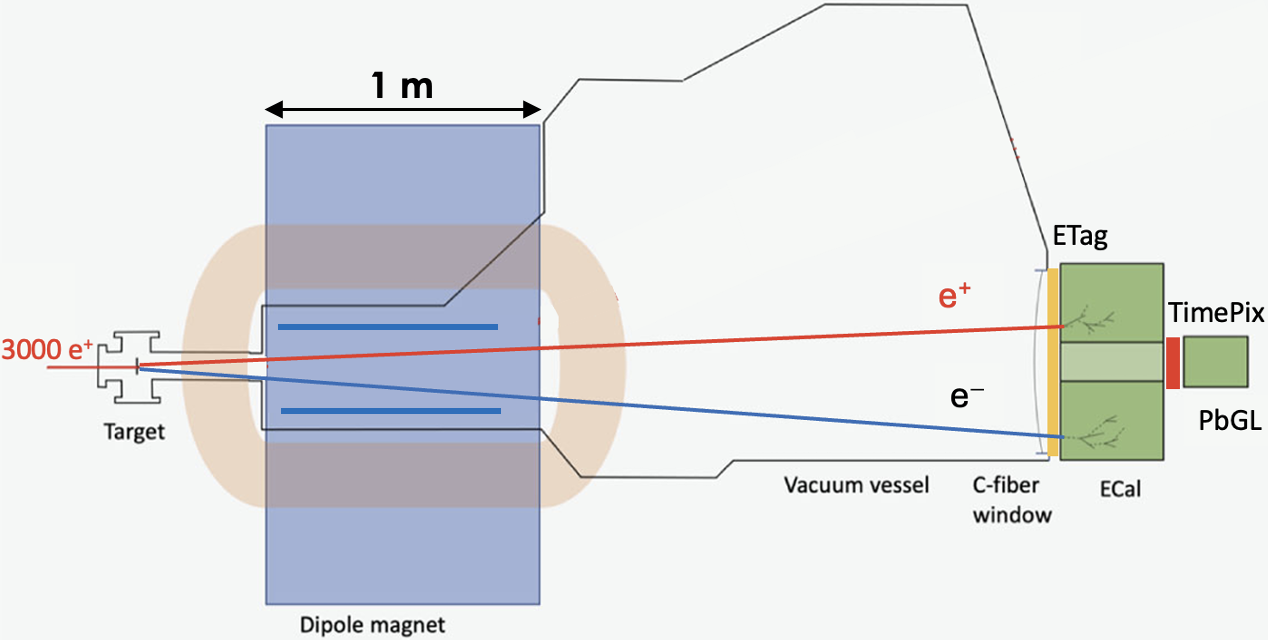}
    \caption{Top-view layout of the PADME detector in the Run III configuration.} 
    \label{fig:padmedetector}
\end{figure}

A detailed description of the PADME detector in Run III, sketched in figure~\ref{fig:padmedetector},  is provided in ref.~\cite{Bertelli:2024ezd}. Single detector performance studies are detailed in ref.~\cite{PADME:2022fuc}. Here we provide a high-level overview of the main detector components relevant to the Run III analysis.

The positron beam is made to interact with an active polycrystalline diamond target operating in a $10^{-6}\unit{mbar}$ vacuum~\cite{PADME:2022fuc}. The target has transverse dimensions of $20\times20\unit{mm}^2$ and a nominal thickness of $\sim100\micron$. Readout strips are etched from the upstream and downstream faces of the target via LASER-induced graphitization. Each strip has a width of 850\micron and pitch of 1\mm. The upstream face provides the vertical ($Y$) and the downstream the horizontal ($X$) beam coordinate~\cite{Oliva:2019alx}. A beam spot with dimensions around $1\times1\unit{mm}^2$ incident on the target is maintained throughout data taking. The stability of the beam is assured by both online and offline monitoring using the charge barycenters in $X$ and $Y$ averaged over one minute of data taking.

The particles resulting from the beam-target interaction, as well as the non-interacting beam, proceed downstream in the vacuum chamber, which fits within the $112\times23\unit{mm}^2$ gap of a dipole magnet. The dipole magnet is turned off during the entire Run III data taking. The vacuum chamber is approximately 3.5\unit{m} long. The downstream end is sealed with a carbon fiber flange of thickness 2.5\mm and diameter 889\mm, which is located immediately upstream of the main detectors.

The electromagnetic calorimeter (ECal) is a matrix of 616 Bismuth Germanate (BGO) crystals refurbished from CERN's L3 Experiment at LEP and read out by photomultiplier tubes~\cite{Albicocco:2020vcy}. Each crystal has transverse dimensions of $21\times21\unit{mm}^2$ and length of 230\mm, corresponding to more than 20 radiation lengths. A central square region with dimensions of $5\times5$ crystals, referred to as the ECal hole, is not instrumented to allow for the passage of non-interacting beam particles and photons emitted by Bremsstrahlung radiation.  The ECal transverse sensitivity region therefore spans the radial range 52.5--300\mm, which, considering the distance from the target of approximately 3.5\unit{m}, corresponds to a laboratory angular emission range of 14--82\unit{mrad}. The ECal signals are digitized in a 1\mus window with rate 1\unit{GS/s}. The event reconstruction then identifies true scintillation signals, determines the timestamp of the signal peak, and estimates the  charge of the signal to form digitized ``hits''. The ECal response is calibrated using cosmic rays before the beginning of Run III~\cite{Albicocco:2020vcy}. The gain of each channel is continuously monitored thanks to a cosmic-ray trigger operating in parallel with the beam data taking. The absolute energy calibration relies on positron interactions.

Immediately upstream of the ECal sits a scintillator-based hodoscope referred to as the ETag, which is read out with silicon photomultipliers (SiPMs). The ETag was installed in Run III to provide partial $e^\pm$/$\gamma$ separation capability. The ETag is not used in the analysis presented in this paper.

A high-granularity silicon pixel detector, the TimePix, is used to monitor the beam parameters at the downstream end of the experimental setup~\cite{timepix:padme-beam}. The system comprises a $2\times6$ array of Timepix3 sensors~\cite{Poikela_2014}, covering an active area of $28\times84\unit{mm}^2$. 
Each Timepix3 chip contains $256\times256$ pixels, 
each with transverse dimensions of $55\times55\micron^2$, with a 300\micron-thick silicon layer. 
The TimePix detector was continuously available for approximately half of the Run III duration. When unavailable, the beam direction at the target is derived from the center of gravity (\cog) between the two particles in the final state. When both estimates are available, the two evaluations have been verified to be mutually compatible.

Finally, an SF57 lead-glass block taken from the calorimeter of the OPAL Experiment at CERN was installed to provide a precise measurement of the positron flux. The block is read out by a 2-inch 
photomultiplier tube, optically connected via a cylindrical SF57 light guide. The photomultiplier divider had been previously refurbished for use in the large angle veto system of the NA62 Experiment~\cite{NA62:2017rwk}. The block was positioned downstream of the ECal hole and of the Timepix assembly, thus absorbing all beam particles and providing a measurement of the total energy deposited by the beam, per bunch.

\section{Run III analysis strategy}
\label{sec:Chap3}

The primary analysis observable, \longgR, is defined as the ratio between the number of observed events with a two-body final state, \Ntwo, and the expected number of background events. The latter is computed by multiplying the measured number of positrons on target (\pot) per bunch, \npot, by the expected number of SM background events per \pot, \B:

\begin{equation}
    \longgR = \frac{\longNtwo}{ \longnpot\times \longB}.
    \label{eq:GR}
\end{equation}

All quantities in eq.~(\ref{eq:GR}) are evaluated for given values of the \com energy, \sqrts. In the absence of a signal, \gR compares the total observed cross section for events with a two-body final state to the corresponding SM cross section, so the expectation is that $\longgR=1$. In the presence of a signal, however, this expectation is modified in the following way:

\begin{equation}
    \longgR =\left[1+\frac{\longS\longeps}{\longB}\right],
    \label{eq:GRSignal}
\end{equation}

\noindent where \S is the signal yield per \pot and \eps is the combination of signal selection efficiency and geometric acceptance. The use of the ratio \longeps/\longB significantly reduces detector-related systematic uncertainties, as both signal and background processes produce a nearly identical detector illumination.

Several theoretical and experimental effects may induce deviations from unity in the \longgR ratio. These are expected to be linearly dependent on \sqrts and are collectively referred to as \longK. Thus, in the background-only (B-only) scenario, the expectation is modified to

\begin{equation}
    \longgR = \longK,
    \label{eq:GRBkgSlope}
\end{equation}

\noindent and in the signal-plus-background (S+B) scenario, it becomes

\begin{equation}
    \longgR = \longK\left[1+\frac{\longS\longeps}{\longB}\right].
    \label{eq:GRSignalSlope}
\end{equation}

\Ntwo and \npot are measured from data, while \B and $\eps/\B$ are determined from Monte Carlo (MC) simulation. The expected signal yield per positron, \S, is estimated using both theoretical and experimental information. Finally, the expected values for the constant term and the \com-energy slope term of \K are determined from data. As detailed in section~\ref{sec:statTreat}, the ultimate goal of the analysis is a statistical comparison between the B-only hypothesis, eq.~(\ref{eq:GRBkgSlope}), and the S+B hypothesis, eq.~(\ref{eq:GRSignalSlope}).  

\subsection{The Run III dataset}

\begin{figure}[t]
    \centering
       \centering
    \includegraphics[width=0.55\textwidth]{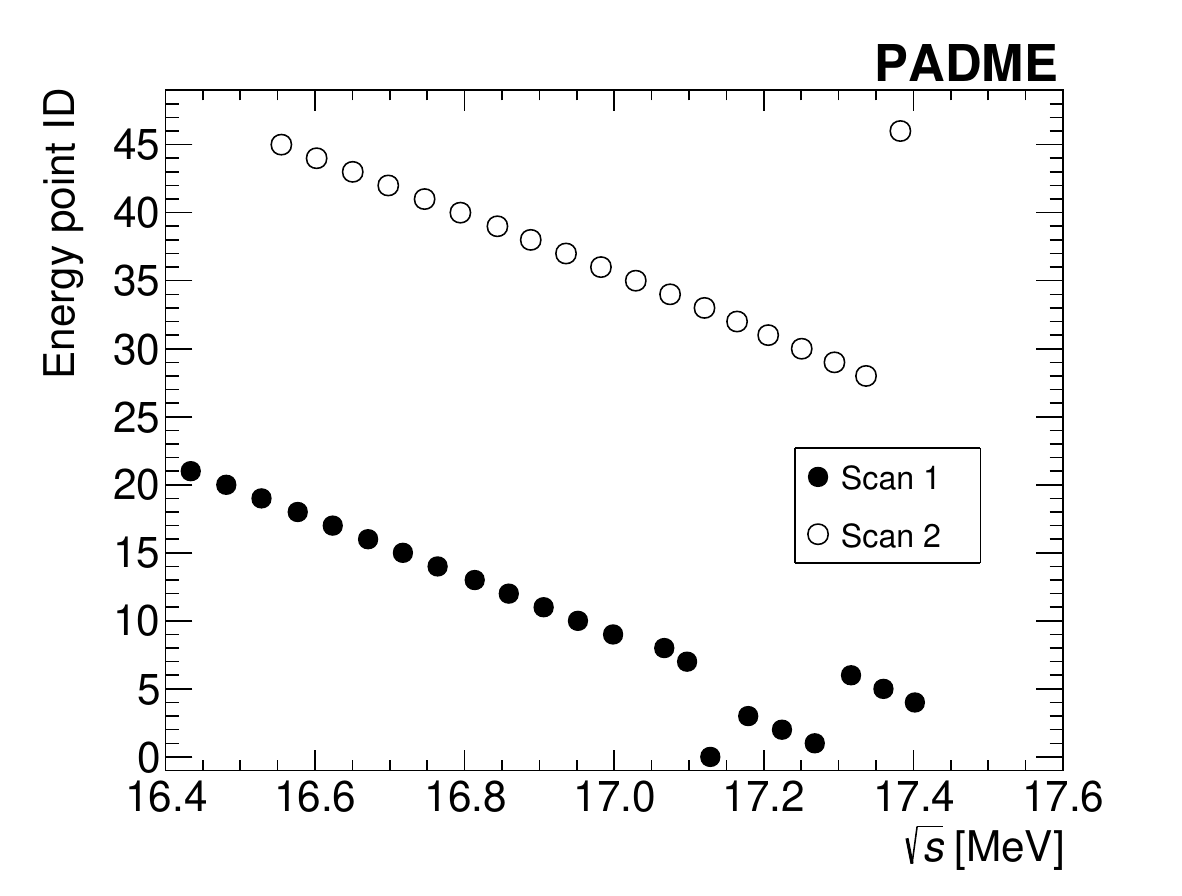}
        \caption{Chronological ID of the energy points taken during the resonance scan \vs \com energy. Filled (open) points refer to Scan~1 (Scan~2). 
        }
        \label{fig:Scans}
    \hspace{0.05\textwidth}
    \end{figure}

The data-taking effort is divided into runs, which are the smallest block periods of continuous data acquisition with stable nominal beam conditions (beam optics, and average beam position, angle, and energy), usually lasting around 8 hours each. The ECal energy scale is evaluated at the run level. A few (usually three) runs with the same nominal beam energy are grouped into an energy point. The beam energy scan is performed by varying the current of the BTF-line dipole elements by an amount corresponding to twice the beam energy spread; with a relative spread of 0.25\%, the energy steps correspond to approximately 1.5\MeV. A first scan (Scan~1) is performed with the nominal beam energy ranging from 295.8 to 262.2\MeV, in decreasing steps. A second scan (Scan~2) covers energy points ranging from 295.1 to 267.7\MeV. The energy points of Scan~1 and Scan~2 are interspersed, so in total all points are spaced out by approximately 0.75\MeV, corresponding to the beam energy spread. Scan~1 and Scan~2 lasted 6 weeks each. Scan points adjacent in energy are therefore acquired with different detector and beam conditions, which provides an additional systematic cross-check. The distribution of energy points, sorted chronologically, is shown in figure~\ref{fig:Scans}.

\begin{figure}[t]
  \centering \includegraphics[width=0.55\textwidth]{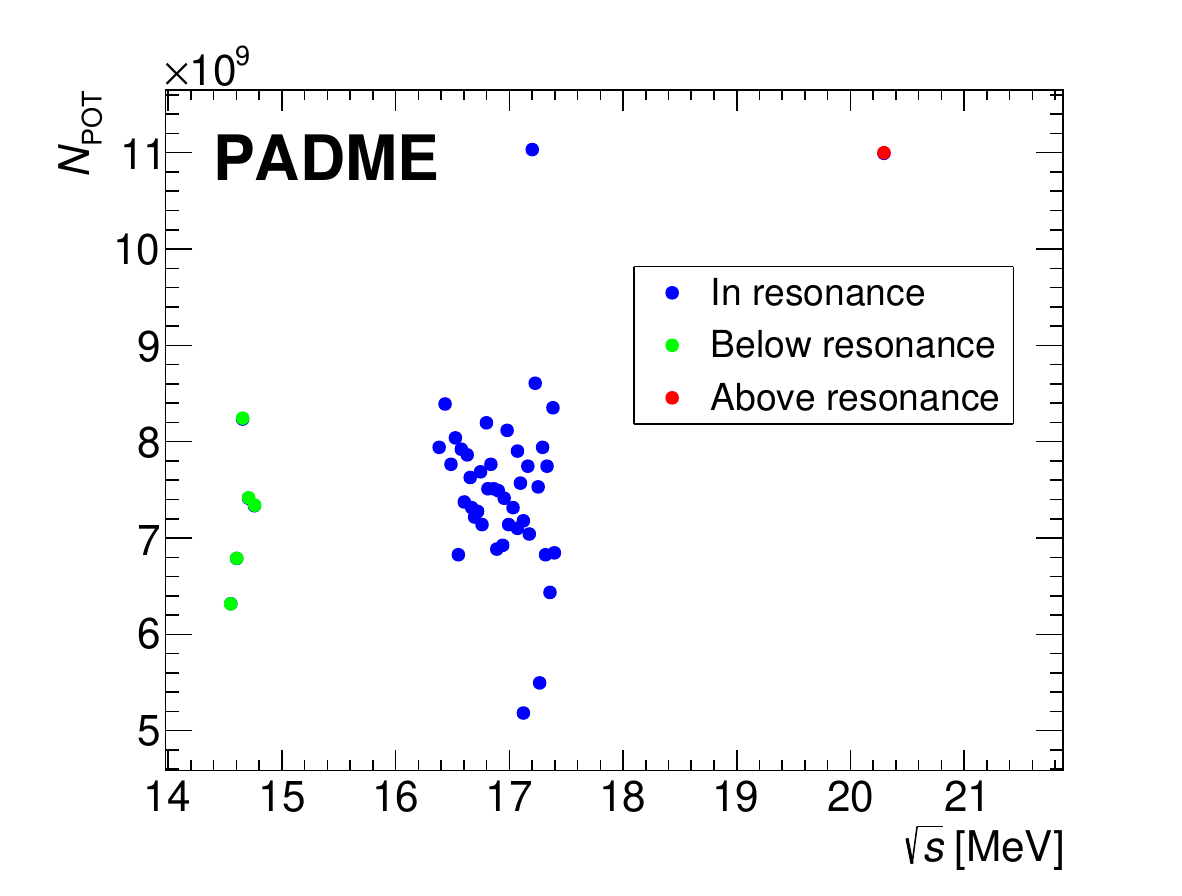}
         \caption{Distribution of \npot \vs \com energy. Blue points represent the scan region, and green and red ones the out-of-resonance data. The latter are used for the \npot calibration and for scaling the beam flux. The \npot variation is due to fluctuations in data-taking efficiency and in machine uptime.}
        \label{fig:padmedataset}
\end{figure}

Two additional datasets are collected: one above the resonance region, with a beam energy of 402\MeV ($\sqrts=20.28\MeV$), and another below, including five energies from 205 to 211\MeV (\sqrts~from 14.5 to 14.7\MeV). Both out-of-resonance samples are immune to contributions from the X17 resonance and therefore serve as useful normalization and background study tools. The number of collected \npot \vs \sqrts is shown in figure~\ref{fig:padmedataset}. 


The analysis of each energy point relies on a dedicated MC simulation of the detector response~\cite{Leonardi:2017lxh} based on the \GEANTfour framework~\cite{GEANT4:2002zbu}. The simulation includes the proper beam parameters (energy, position, angle, and transverse spread at the target and at the ECal) and a detailed detector description (\eg, dead and noisy ECal cells).

\subsection{Event selection}\label{sec:signalsel}

Events of interest are selected from the ECal information. A 1\MeV threshold is applied in order for an ECal hit to be reconstructed. Hits nearby in time (within 6\unit{ns}, corresponding to 3$\sigma$ deviations) and space (within two ECal cells) are grouped to form energy clusters. A cluster is identified if at least one hit features a reconstructed energy above 15\MeV. This hit becomes the seed crystal. As expected, both hit and seed energy thresholds are found to lead to a very loose selection of the ECal clusters forming the \Ntwo observable. The cluster time and position are determined from energy-weighted averages of the cluster hits.

\begin{figure}[t]
    \centering
\includegraphics[width=0.65\linewidth]{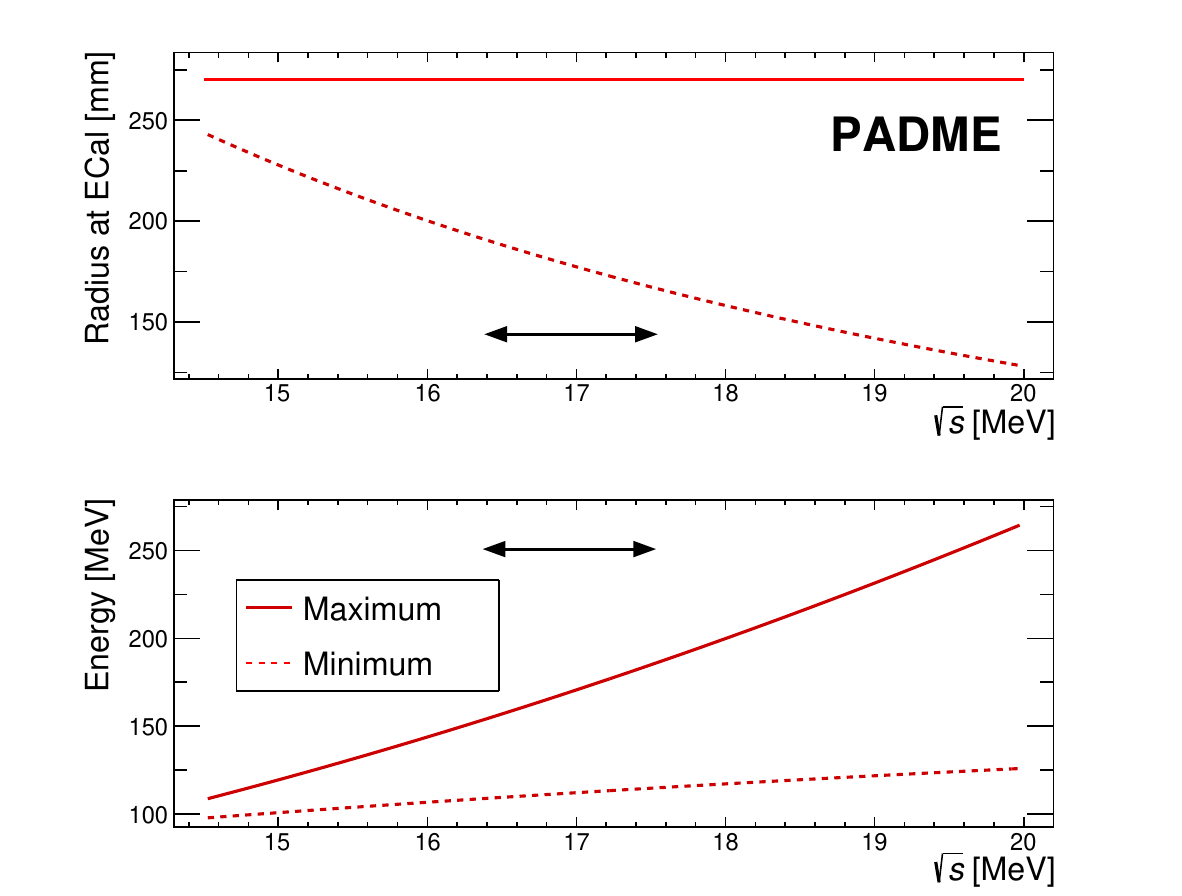}
    \caption{Radial positions (top) and energies (bottom) of the final-state particles allowed by the geometric acceptance of the ECal and the event selection, shown as a function of \com energy. Solid lines denote the maximum and dashed lines the minimum values. The maximum radial position is imposed as a constant value (270\mm) to avoid edge effects. The double-headed arrow in each panel shows the range used for the energy scan in the analysis.
    }
    \label{fig:kinematics}
\end{figure}

Events with a two-body final state are identified by the existence of two spatially separated, in-time energy clusters. The time difference between the clusters in the pair is required to be within ${\pm}5\unit{ns}$, which corresponds to a $3\sigma$ window~\cite{PADME:2022fuc}.
The transverse radial cluster position relative to the ECal geometric center, $R$, is naturally limited by the ECal dimensions to be below 300\mm. Therefore, requiring that both interaction products ($e^+e^-$ or $\gamma\gamma$) be within the ECal geometric acceptance implies a minimum value for the radial position and a limited energy range for each cluster. Both limits depend on the \com energy. To avoid edge effects, we require a maximum radius $\Rmax=270\mm$, which corresponds to a distance from the edge of approximately 1.5 times the size of each ECal cell. The corresponding energy range and minimum radius (\Rmin) are shown in figure~\ref{fig:kinematics}. 

The selection algorithm applied to candidate in-time cluster pairs includes several fine tunings:

\begin{itemize}

\item The variation of the transverse beam position at the ECal during all of Run III is on the order of 10\mm~\cite{Bertelli:2024ezd}. The beam position at the ECal is determined with an uncertainty below the mm level for each run. Events with a two-body final state are selected using a spatially unconstrained set of in-time cluster pairs; the \cog position is evaluated as the energy-weighted barycenter of the two clusters. The uncertainty in this estimate is also below the mm level, which is determined by fitting the \cog position and the TimePix illumination response with multivariate normal distributions and then comparing the results for the $X$ and $Y$ mean values~\cite{Bertelli:2024ezd}. To remain as independent as possible from  variations in the beam position, the ECal kinematic center  used to evaluate the transverse position of clusters is fixed to the \cog position.

\item The beam spot at the ECal has dimensions of around 7--10 and 11--12\mm in the $X$ and $Y$ directions, respectively, depending on the beam energy and optics. To be as independent as possible of the beam spot size and of the spatial resolution of cluster reconstruction, a safety margin is applied: $\Rmin \to \Rmin - D$, where $D = 31.5\unit{mm}$ corresponds to 1.5 times the ECal cell size;

\item A residual magnetic field of 12.5\unit{G} was present in the PADME analyzing magnet, as measured with a Hall probe, and is included in the MC simulation~\cite{PADME:2022ysa}. Nevertheless, to account for possible systematic uncertainties in the related bending of the particles in the $e^+e^-$ final state, the more stringent condition $\Rmin < R < \Rmax$ is applied to the highest energy cluster in the candidate pair, while the weaker condition $R > \Rmin$ is applied to the lowest energy one.

\item The dimensions of the magnet aperture limit the geometric acceptance in the $Y$ coordinate. This loss is found to be asymmetric between the positive and negative $Y$ directions. To avoid systematic effects arising from this asymmetry, we do not consider cluster pairs with laboratory azimuthal angles $\phi_{1,2}$ in the range $|\phi_{1,2}-\pi/2|<\phi_0$ or $|\phi_{1,2}-3\pi/2|<\phi_0$, where $\phi_0\simeq0.87$~rad. The acceptance loss induced by this requirement is about 30\%. 

\end{itemize}

\begin{figure}[t]
\centering
\includegraphics[width=0.55\textwidth]{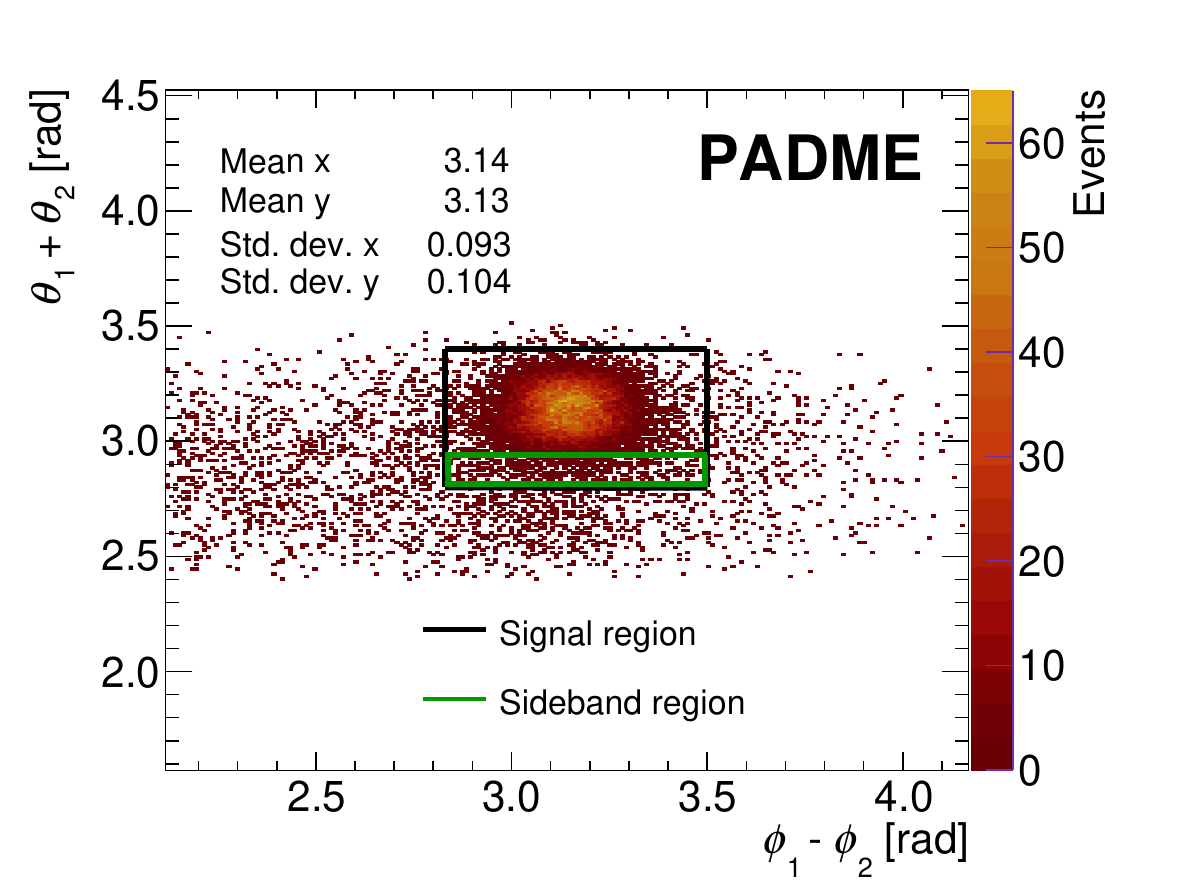}
    \caption{Distribution of candidate cluster pair opening angles in the \com frame: $\theta_1+\theta_2$ \vs $\phi_1-\phi_2$. Signal and sideband regions are indicated by the black and green lines, respectively.} 
    \label{fig:openingCoM}
\end{figure}

For all cluster pairs, 
the laboratory frame momentum vector of each particle in the pair is evaluated by connecting the beam spot position at the target to the cluster position at the ECal. Given the two particle directions in the laboratory, the opening angle in the \com frame is estimated by Lorentz-boosting the corresponding laboratory frame quantities. For $e^+e^-$ final states, by neglecting terms of the order of the electron mass to particle energy ratio, the opening angle in the \com frame can be evaluated independently of the cluster energies $E_{1,2}$. For $\gamma\gamma$ final states, no approximation is needed. The distribution of events across the sum of the polar angles and the difference of the azimuthal angles in the \com, $\theta_1+\theta_2$ and $\phi_1-\phi_2$, respectively, is shown in figure~\ref{fig:openingCoM}. The signal region centers around $\theta_1+\theta_2\simeq\pi$ and $\phi_1-\phi_2\simeq\pi$. The bending induced by the residual field in the PADME dipole magnet results in asymmetric broadening of the $\theta_1+\theta_2$ distribution.

The signal region is defined by the requirements $2.9 < \theta_1+\theta_2 < 3.4\unit{rad}$ and $2.83<\phi_1-\phi_2<3.5\unit{rad}$. A sideband region defined as $2.8 < \theta_1+\theta_2 < 2.9\unit{rad}$ and $2.83<\phi_1-\phi_2<3.5\unit{rad}$ is used to scale the simulated background to the one measured in data. A 4\% background component, originating from Bremsstrahlung radiation in the target, is subtracted to obtain the final \Ntwo yield.
\section{Analysis-level corrections} 
\begin{figure}[t]
    \centering
\includegraphics[width=0.55\linewidth]{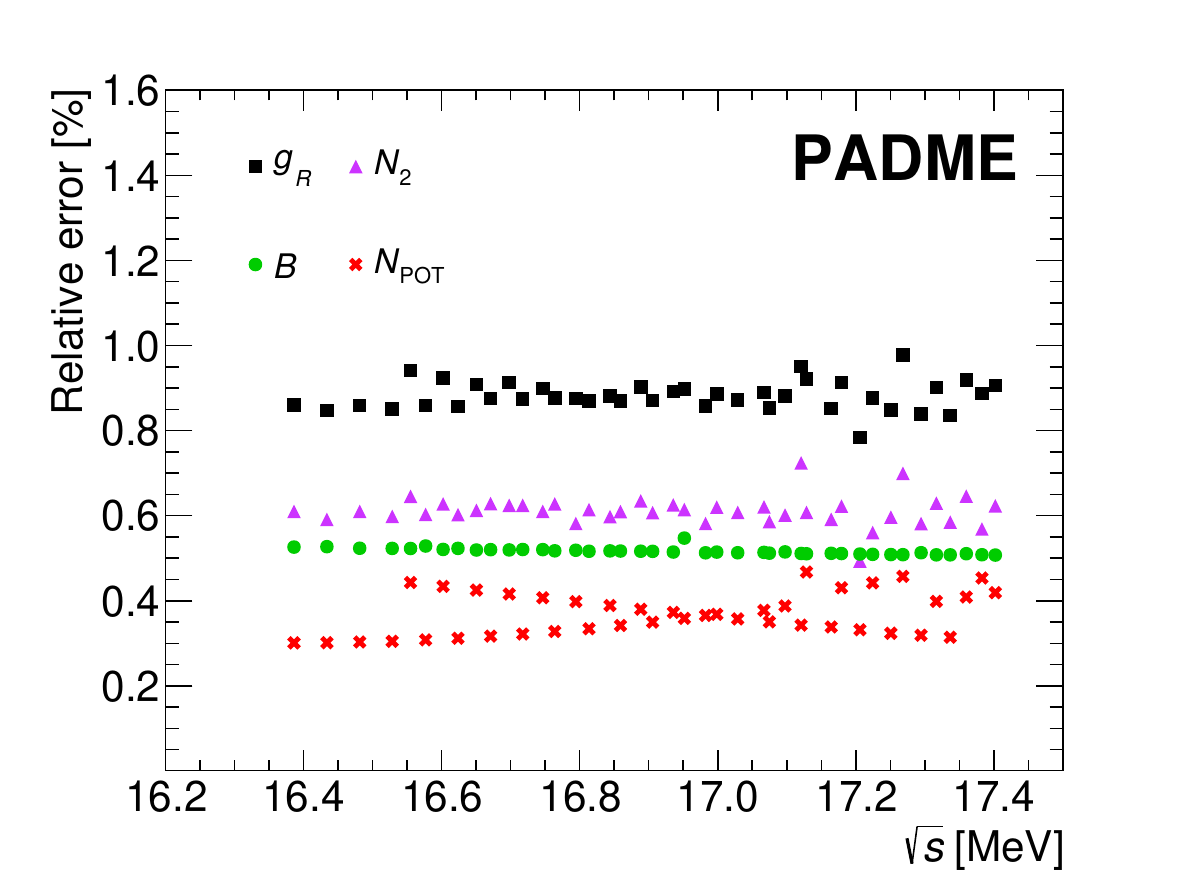}
    \caption{Relative error \vs \sqrts for sources of uncorrelated uncertainty affecting the \gR observable.}
    \label{fig:uncorrelatederr}
\end{figure}

%
The procedure for estimating the quantities entering eq.~(\ref{eq:GRSignalSlope}) is discussed here. Sources of statistical and systematic uncertainty are divided into contributions that are either correlated or uncorrelated across energy points in the scan. The sources of uncertainty and their impact on the final result are summarized in table~\ref{tab:errorbudget}. Typical values are reported for uncorrelated errors. In figure~\ref{fig:uncorrelatederr}, the relative uncertainties are shown as a function of \com energy.

\begin{table}[t]
    \centering
    \renewcommand{\arraystretch}{1.2}
    \begin{tabular}{lcc}
\multicolumn{3}{c}{Uncorrelated errors} \\
    [\BigSkip]
Source & Uncertainty (\% per energy point) & Remarks \\  \hline 
\longNtwo & 0.60 & Statistical  \\  
\longB & 0.54 & \ref{sec:Bexp} \\ 
\longnpot & 0.35 & \ref{sec:ageing} \\  \hline
Total on \longgR & 0.88 & \\ 
    \multicolumn{3}{c}{\longK, constant term}
    \\
    [\BigSkip]
    Source & Uncertainty (\%) & Remarks \\   \hline     
Lead-glass calibration & $2.0$ & ref.~\cite{Bertelli:2024ezd} \\
Absolute \B yield & $1.8$ & \ref{sec:yield} \\
Energy-loss correction to \npot & $0.5$ & \ref{sec:leakage} \\
Radiation-induced correction to \npot & $0.3$ & \ref{sec:ageing} \\   \hline
Total & $2.8$ &\\
    \multicolumn{3}{c}{\longK, slope term}
    \\
    [\BigSkip]
    Source & Expected value ($\%/$MeV) & Remarks \\   \hline     
    Radiative corrections & $-0.6\pm0.2\pm0.6$ & \ref{sec:yield} \\
  \hline
    Total & $-0.6\pm0.6$ & \\ 
    \end{tabular}
    \caption{Top: typical uncertainty values for sources that are uncorrelated across energy points. Middle: uncertainty of the constant term in \longK. Bottom: expected values of the slope term in \longK.\label{tab:errorbudget}}
\end{table}

\subsection{Determination of \texorpdfstring{\B}{B}\label{sec:Bexp}}
The expected number of background SM events and the signal selection efficiency are determined from MC simulation. Various sources of systematic uncertainty were investigated, the most relevant ones stemming from the reliability of the simulation, the impact of beam parameter variations, and the uncertainty in the production rate. A summary of the various sources of 
uncorrelated uncertainty in the estimate of \longB is given in table~\ref{tab:berrorbudget}.

\begin{table}[t]
    \centering
    \renewcommand{\arraystretch}{1.2}
    \begin{tabular}{lcc}
\multicolumn{3}{c}{\longB uncorrelated error}  \\
    [\BigSkip]
Source & Uncertainty (\% per energy point) & Remarks \\ \hline     
MC statistics & 0.40 & \NA \\ 
Tag and probe efficiency & 0.35 & \ref{sec:tagandprobe} \\ 
Cut stability & 0.04 & \ref{sec:cutstability} \\
Beam spot variations & 0.05 & \ref{sec:cutstability}\\
\hline
Total on \longB & 0.54 & \\  
\end{tabular}
    \caption{Sources of uncorrelated uncertainty in \longB.\label{tab:berrorbudget}}
\end{table}

%

\subsubsection{Reconstruction efficiency}
\label{sec:tagandprobe}

The MC simulation~\cite{Leonardi:2017lxh} features a detailed \GEANTfour~\cite{GEANT4:2002zbu} representation of the ECal geometry and structure. The same hit- and seed-level thresholds, as well as clustering selections, are applied in simulation and in data. A smearing of the time and energy response function of the ECal is applied to better match the data, both at the hit and cluster levels. 

\begin{figure}[t]
    \centering    \includegraphics[width=0.55\linewidth]{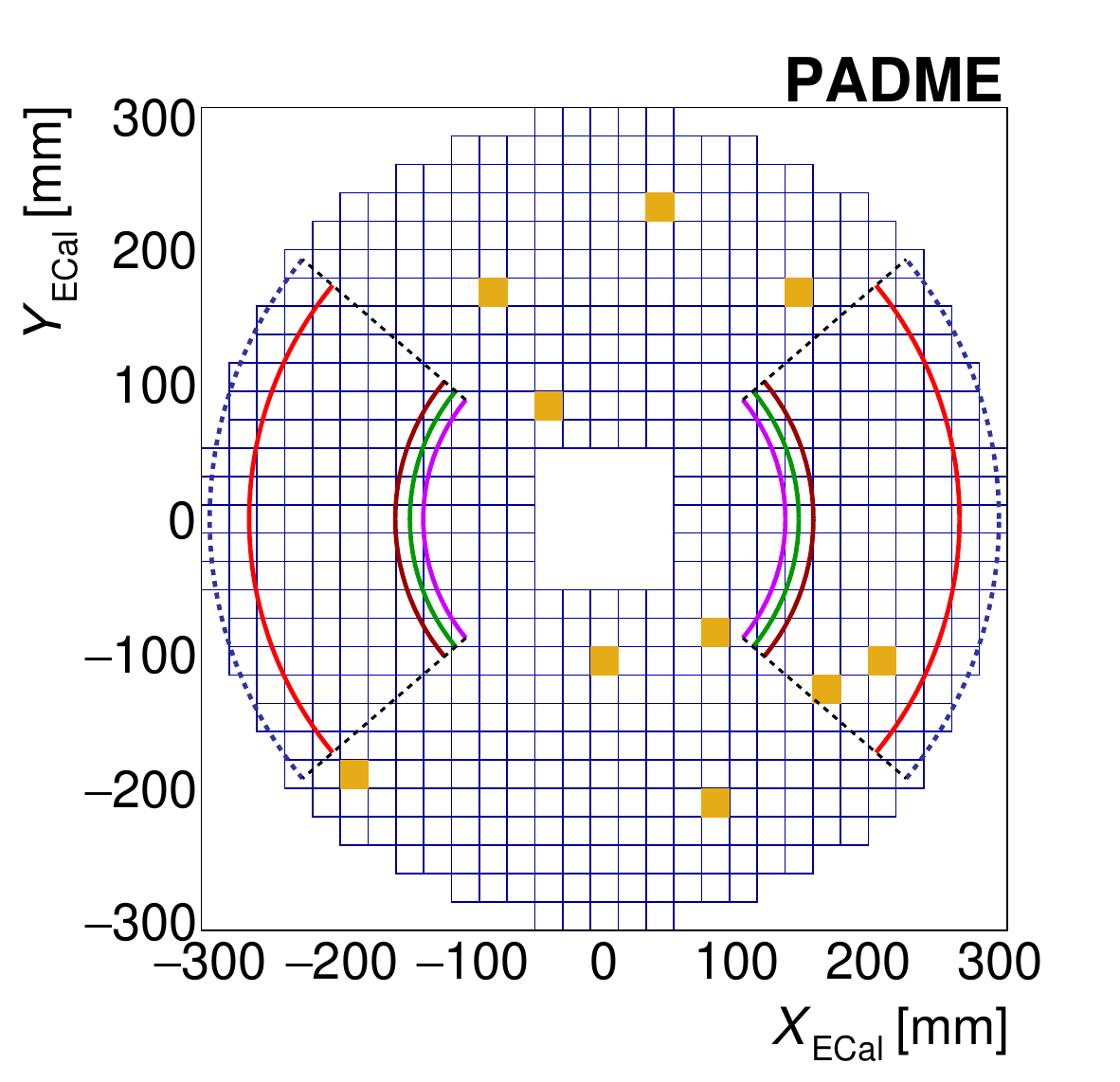}
    \caption{Schematic illustration of the ECal acceptance in the analysis. Events with two-body final states can be considered for selection only if the two particles are found within the two regions bound by the black dashed lines in the radial direction and by the colored arc segments in the azimuthal direction. The outermost solid bright red (dashed blue) arc segment depicts the nominal maximum radial position used in the analysis for the cluster with the highest (lowest) energy. The remaining inner segments depict several minimum radial positions for both clusters, which vary according to \sqrts. The brown, green, and pink segments correspond to \sqrts values of 16.4, 16.9, and 17.4\MeV, respectively. The ECal cells flagged as dead or noisy are shown as filled orange squares.}
    \label{fig:ECalYX}
\end{figure}

Dead and noisy ECal cells were identified from data occupancy distributions. Ten of 616 cells were flagged as dead or noisy in at least one run. Such cells were masked in the reconstruction of the entire dataset, both in data and in simulation. Two dead cells were included in the region accepted for the selection of events with two-body final states, as shown in figure~\ref{fig:ECalYX}.

The reliability of the simulation-based efficiency estimate is checked for each energy point using the tag-and-probe method. Events with two-body final states are selected by identifying a single ECal cluster (tag), and the fraction of such events where a second cluster (probe) is found stands as a proxy for the ECal reconstruction efficiency. Tags are selected as follows:

\begin{enumerate}
    \item The ECal tag cluster must be found within the \sqrts-dependent radius range \Rmin--\Rmax. The radius is evaluated with respect to the average \cog, similarly to the signal selection, as described in section~\ref{sec:signalsel};
    \item The position of the tag cluster and the expected position of the probe must be in the same region defined by the azimuthal angles of the signal selection, as described in section~\ref{sec:signalsel};
    \item The difference $\Delta E$ between the energy of the tag cluster and its expected energy given the reconstructed position must be within ${\pm}15\MeV$.
\end{enumerate}


A significant number of background events populate the sample of tag clusters because of accidental beam activity and Bremsstrahlung radiation from beam particles incident on the target. This background is subtracted based on the $\Delta E$ distribution of signal and background templates. The accidental background templates are derived from data taken with the target physically removed from the line of sight of the beam. Probes are then selected as follows:

\begin{enumerate}
\item A probe cluster must be found within a ${\pm}5\unit{ns}$ time window relative to the tag; 
\item The transverse distance between the tag and probe clusters must be larger than 60\mm to avoid split clusters;
\item The sum of the polar angles and the difference of the azimuthal angles in the \com frame must be in the same regions used for the signal selection.
\end{enumerate}

\begin{figure}[t]
    \centering
    \includegraphics[width=0.55\linewidth]{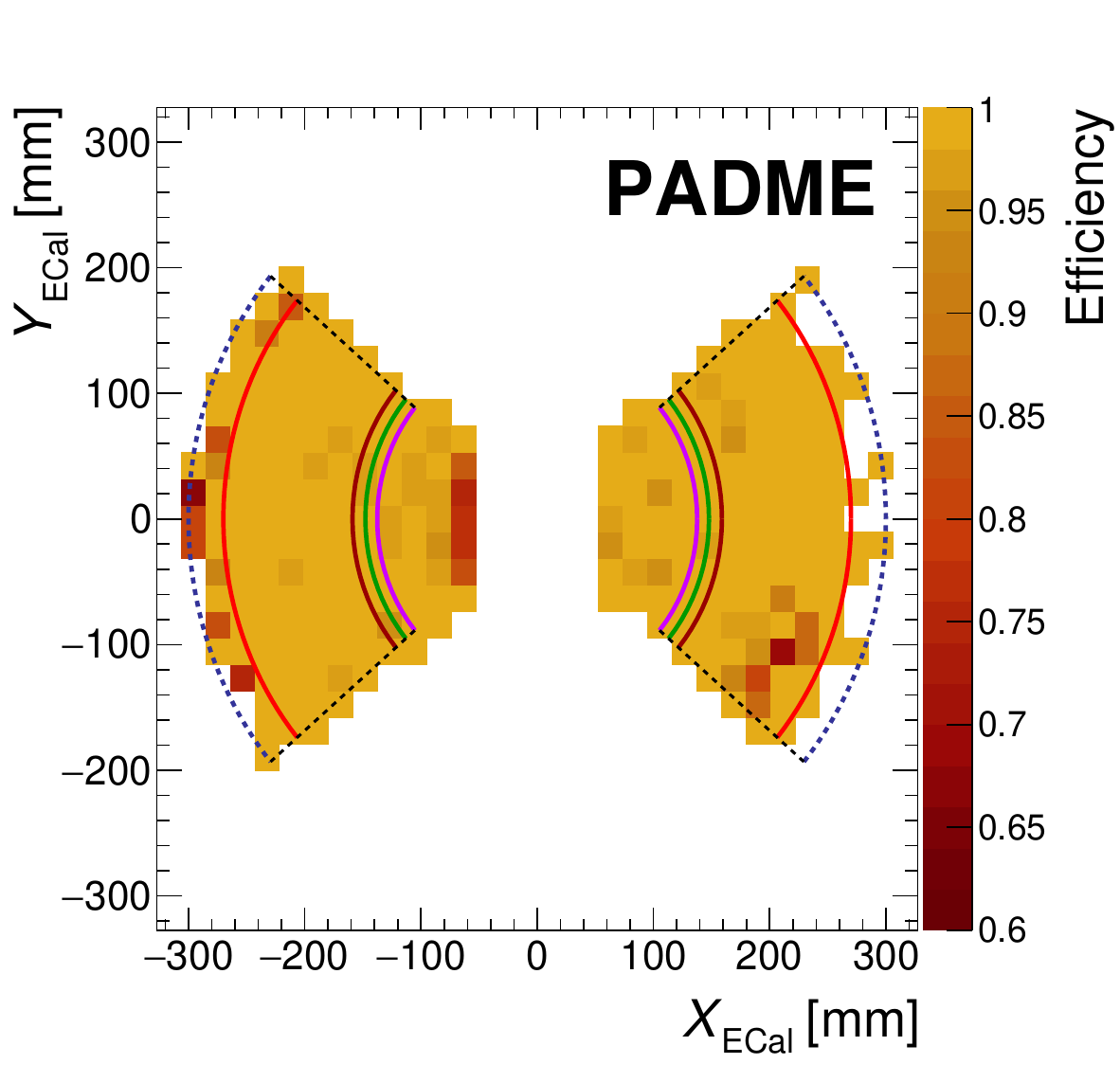}
    \caption{True single-particle reconstruction efficiency as a function of impact point on the ECal, measured with simulation and including both $e^+e^-$ and $\gamma\gamma$ final states. The accepted regions are shown as the sectors bound by the dashed radial lines and azimuthal arc segments. The outermost solid bright red (blue dashed) arc segment depicts the maximum radial position for the cluster with maximum (minimum) energy. The remaining inner segments depict several minimum radial positions, which vary according to \sqrts. The brown, green, and pink segments correspond to \sqrts values of 16.4, 16.9, and 17.4\MeV, respectively. The transverse beam spot for this particular energy point in the scan has position $\{X_\mathrm{ECal},Y_\mathrm{ECal}\} = \{-12.4,-0.5\}\mm$.}
    \label{fig:TrueEff}
\end{figure}

The ratio of probe-matched events to background-subtracted tag events, defined as the ECal reconstruction efficiency, is then evaluated as a function of the expected probe energy. This tag-and-probe efficiency is a proxy for the true ECal reconstruction efficiency, defined as the fraction of events where a cluster is reconstructed given that a final-state $e^{\pm}/\gamma$ was emitted after $e^+e^-$ annihilation occurred at the target. From simulation, the true efficiency for selected events is expected to be unity with less than 1\% uncertainty, except in specific regions of the ECal, as shown in figure~\ref{fig:TrueEff}, like the outer edges and the ECal crystals adjacent to dead or noisy cells.

The ratio of tag-and-probe efficiency to true ECal efficiency 
is one within a few percent, and is consistent across different energy periods.

\begin{figure}[t]
\centering  \includegraphics[width=0.6\textwidth]{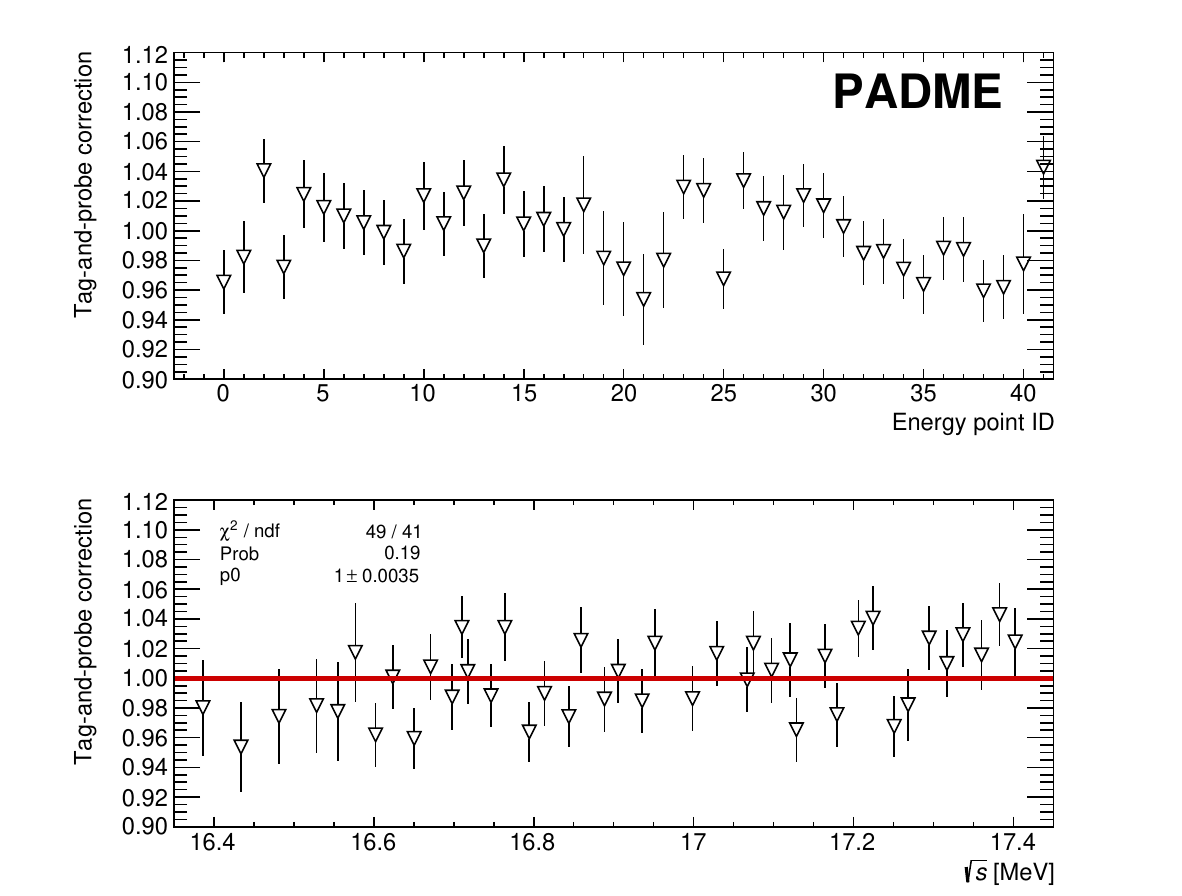}
\caption{Mean value of the energy correction derived with the tag-and-probe method as a function of the chronological ID of the energy point in the scan (top panel) and the \com energy (bottom panel). A fit to a constant is overlaid onto the data points in the bottom panel.} 
\label{fig:tp_effmean}
\end{figure}

The ratio between the data and simulation tag-and-probe efficiencies is found to be independent of \com energy, as shown in figure~\ref{fig:tp_effmean}, where the energy-averaged value of this ratio is plotted for each energy period and as a function of \com energy. A fit to a constant yields a probability of 19\% that the model describes the data and an associated error of 0.35\%, which is included in the list of uncorrelated uncertainties, referred to as the constant term of \longK.





\subsubsection{Reliability of the \texorpdfstring{\B}{B} estimate}

\label{sec:cutstability}

The selection algorithm depends on the expected average beam direction and on the shape of the beam spot at the target, which are determined from the distribution of the deposited charge in the diamond target and from the \cog measurement at the ECal, for events with a two-body final state. Because of the acceptance limitations induced by the magnet bore dimensions, the largest systematic effect arises from vertical beam displacements. Relative acceptance variations range from 0.08\% to 0.1\% per mm of vertical shift, for \sqrts values ranging from 16.4 to 17.4\MeV. An error of 0.5\mm on the beam position from the \cog estimate is taken as conservative as determined from the comparison of TimePix and ECal information. The systematic uncertainty due to beam variations is therefore below 0.05\% for the entire dataset.

A number of effects stemming from imprecise MC simulation modeling could also impact the estimate of \longB and were further considered: 

\begin{itemize}
    \item The reconstruction of clusters in the vicinity of dead or noisy cells;
    \item Multiple Coulomb scattering processes occurring in the passive material traversed by the beam that could lead to acceptance losses;
    \item The reconstruction of clusters corresponding to particles that are incident close to the ECal geometrical edges.
\end{itemize}

\begin{figure}[t]
    \centering
\includegraphics[width=0.6\linewidth]{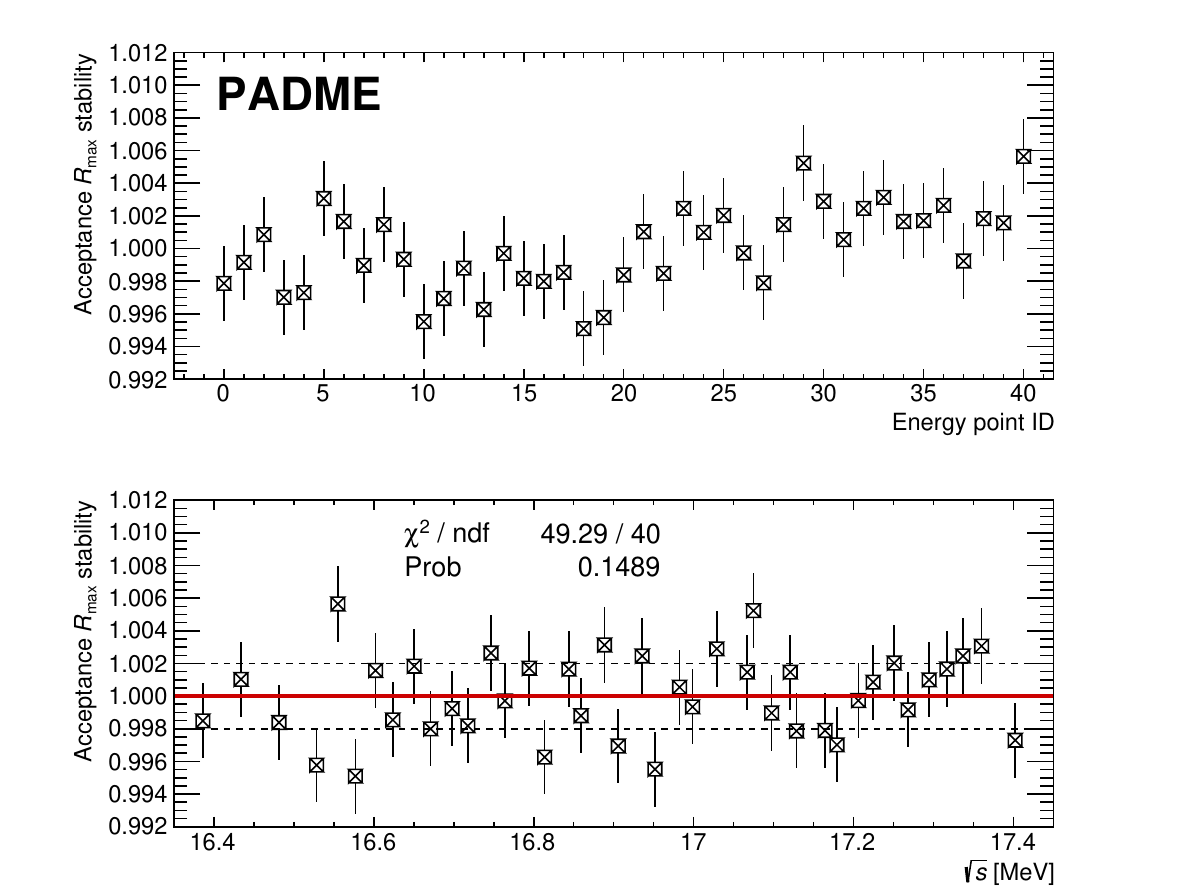}
    \caption{Relative data-to-MC variation of the event yield when shifting the value of \Rmax from 230\mm to 300\mm, as a function of the chronological ID of the energy points in the scan (top panel) and \com energy (bottom panel). In the bottom panel, a fit to a constant is overlaid as a solid red line, while a $1\sigma$ envelope of ${\pm}0.2$\% is shown by the two dashed lines.}
    \label{fig:RMaxStability}
\end{figure}

To investigate these effects, data and simulation expectations for \longB were compared while varying \Rmax from the default value of 270\mm to 230 and 300\mm. 
When going from 230 to 300\mm, the number of accepted events increases by approximately 10\%. The data-to-MC relative variation of the event yield for various energy points is shown in figure~\ref{fig:RMaxStability}. This variation is consistent with a constant number that is independent of the \com energy. A ${\pm}0.2$\% envelope covers approximately 68\% of the energy points. The error on the average of this envelope, 0.035\%, is taken as a source of uncorrelated systematic uncertainty in the \longB estimate.


\subsubsection{Production yield}
\label{sec:yield}

\begin{figure}[t]
    \centering
\includegraphics[width=0.45\linewidth]{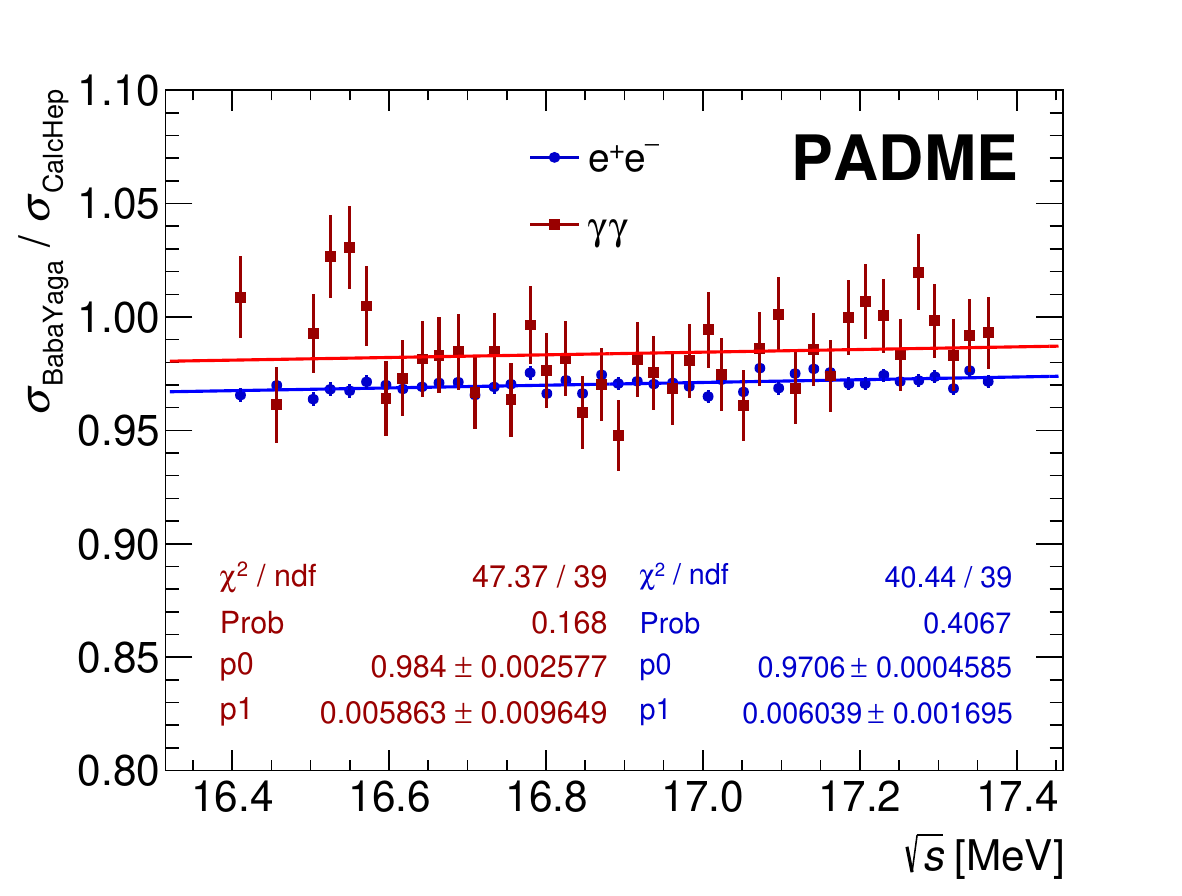}  \includegraphics[width=0.45\linewidth]{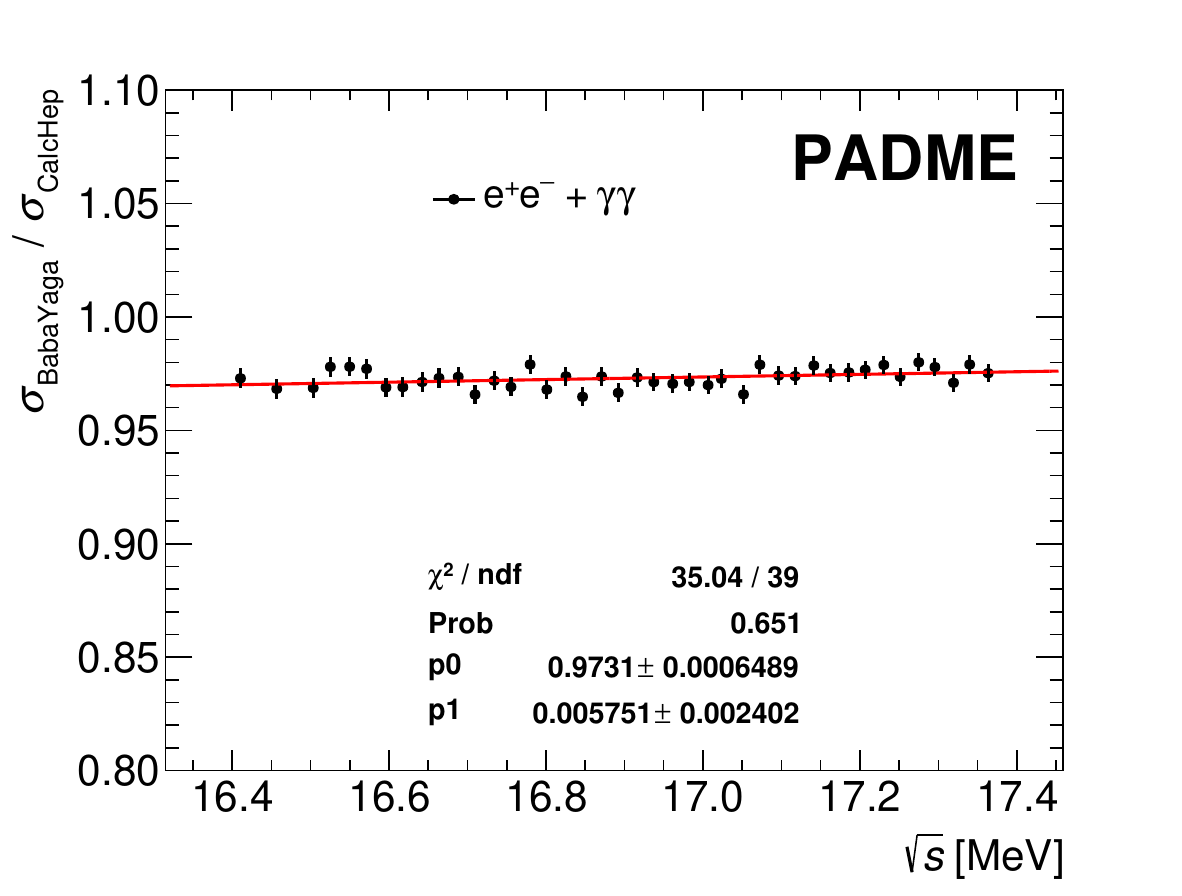}
    \caption{Left: Ratio between the SM cross sections for Bhabha scattering (blue points) and $\gamma\gamma$ production (red points) obtained with the \BabaYaga generator~\cite{Balossini:2006wc,Balossini:2008xr} and the corresponding tree-level evaluations as a function of \sqrts. Right: Ratio between the total SM two-body production cross section from \BabaYaga and the corresponding tree-level evaluation. In both panels, linear fit results are overlaid.}
    \label{fig:radcorr}
\end{figure}

The absolute number of events with a two-body final state is determined from the total cross section for $e^+e^-\to e^+e^-,\gamma\gamma$ scattering and from the product of the target thickness and density, which was previously determined with an uncertainty of around 4\%~\cite{PADME:2022tqr}.
The total effective cross section was determined from MC simulation, including radiative corrections. These were evaluated for both $e^+e^-$ and $\gamma\gamma$ production using the \BabaYaga package~\cite{Balossini:2006wc,Balossini:2008xr}. The ratio of the computed cross section with radiative corrections to the one without is shown as a function of \sqrts in figure~\ref{fig:radcorr}. Including radiative corrections leads to a 3\% decrease in the total cross section at the center of the scan region, $\sqrts=16.92\MeV$, and a \com-energy slope of $0.6\pm0.2\% \MeV^{-1}$. The theoretical uncertainty from \BabaYaga  is claimed to reach 0.1\% at \com energies corresponding to the $\Phi$ meson mass~\cite{Balossini:2006wc}. The allowed angular regions for the final-state particles can significantly affect the expected uncertainty. A conservative evaluation of the expected \sqrts slope is $0.6\%\pm0.6\%\MeV^{-1}$. Events simulated with \BabaYaga were inserted in the \GEANTfour-based MC simulation that includes the detector response. Up to 10 radiative photons can be part of the final state in addition to the main $e^+e^-$ or $\gamma\gamma$ particles of interest. The acceptance is found to be insensitive to radiative corrections.

To minimize sensitivity to the aforementioned uncertainties, the measured yield was normalized to the one obtained from the below-resonance energy points. Several additional systematic uncertainties related to this procedure were considered:

\begin{itemize}
    \item The statistical error in the yield for below-resonance points is 0.4\%;   
    \item The acceptance of the below-resonance points is reduced by more than 30\% with respect to the on-resonance points, with a related systematic uncertainty of 1\%;
    \item The \com-energy slope induced by radiative corrections results in a systematic shift of the yield for below-resonance points, which corresponds to a shift of $(-1.5\pm1.5)\%$ on \longK. This offset is not applied as a correction, since it is covered by the other systematic uncertainties affecting the constant term of \longK.
\end{itemize}

The overall estimated uncertainty on the constant term of \longK is therefore ${\pm}1.8\%$.
\subsection{Determination of \texorpdfstring{\npot}{NPOT}}
\label{sec:NPoT}

The number of \pot per bunch is determined by normalizing the charge collected with the lead-glass calorimeter to the beam energy and then multiplying by a calibration constant. The absolute calibration of the detector was  discussed extensively in ref.~\cite{Bertelli:2024ezd}. The charge corresponding to a single positron at a calibration energy of 402.5\MeV was measured in a dedicated run. The relative uncertainty on the absolute calibration is 2\% and is among the dominant uncertainties on the constant term of \longK. A set of corrections is needed to derive a precise estimate of \npot, as discussed in the following subsections.

\subsubsection{Energy-loss correction}
\label{sec:leakage}

A considerable variation of the beam position and spatial spread is observed in the different energy periods of Run III, as determined from the TimePix and ECal information~\cite{Bertelli:2024ezd}. These effects have two main impacts: i) a variable energy loss in the passive material before entering the lead-glass calorimeter, and ii) a different beam impact point at the lead-glass calorimeter. 
The former is the dominant one and, because of the location of the passive material, is particularly enhanced by beam movements in the vertical direction.

\begin{figure}[t]
    \centering
\includegraphics[width=0.6\textwidth]{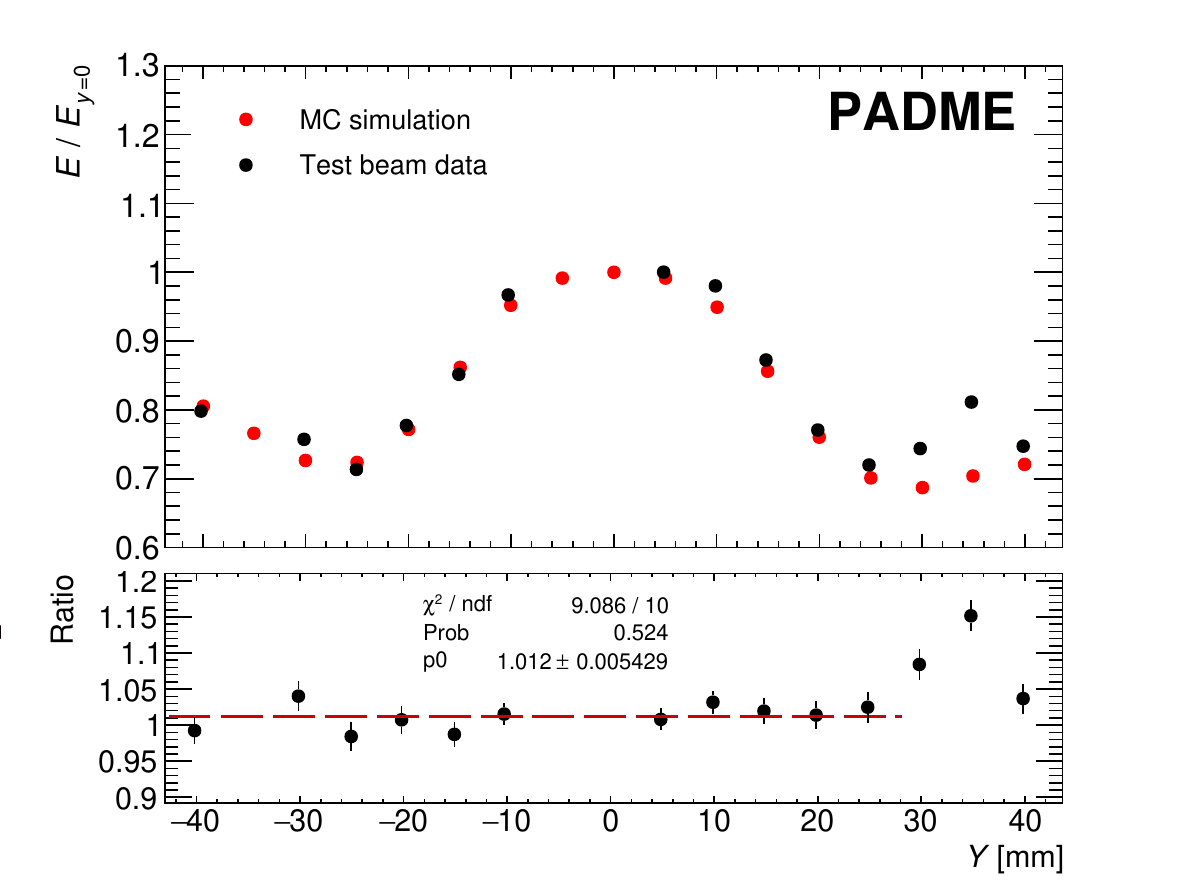}
        \caption{Average energy deposition \vs vertical beam position $Y$ at the front face of the lead-glass calorimeter. Data (black points) are compared to the simulation predictions (red points). The bottom panel displays the ratio between them. A fit to a constant is also overlaid (red dashed line).}
        \label{fig:TB}
\end{figure}

A dedicated test beam campaign was conducted to evaluate and correct for such effects. A horizontal and vertical beam position scan was done using a 285\MeV electron beam with a multiplicity of roughly 500 particles in 10\unit{ns}-long spills. A beam spot of $5\times5\unit{mm}^2$ at the TimePix plane was maintained. The lead-glass calorimeter response relative to the geometrical center was measured and compared to the prediction obtained from a detailed MC simulation, including a faithful mechanical description of the relevant detectors. For the vertical position scan, the data (black dots) are compared to simulation (red dots) in figure~\ref{fig:TB}. The region matching the TimePix sensors, with $\abs{Y}<14\mm$, features the minimum energy loss. 

Particles lose more than 10\% of their energy when traversing the TimePix cooling system in the outer regions. Good agreement is seen between data and simulation throughout the range of vertical positions of interest for Run III, approximately ${-}30$ to 25\mm. The average ratio of data-to-simulation energy deposition contributes a constant correction to \longK, and the related error of ${\pm}0.005$ is also accounted for.

\begin{figure}[t]
  \centering \includegraphics[width=0.6\textwidth]{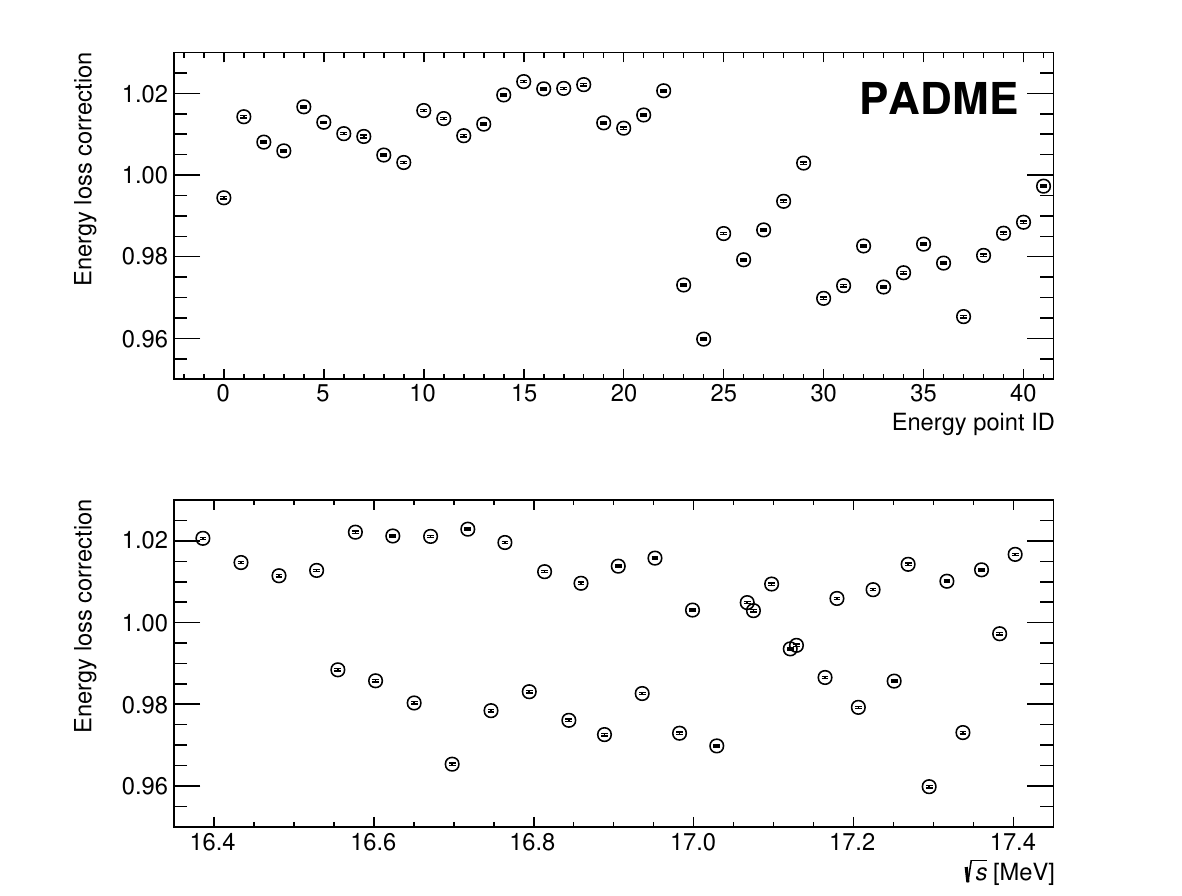}
        \caption{Relative values of the energy-loss correction to \npot as a function of the chronological ID of the energy point (top panel) and the \com energy (bottom panel). The vertical beam impact positions differ significantly between Scan 1 and Scan 2, leading to the vertical shift observed in the top panel, starting with point ID 24.}
    \label{fig:frontal_leakage}
\end{figure}

A MC simulation including a detailed description of the Timepix material budget and of the beam particle interactions is used to evaluate the overall energy loss for the specific beam conditions of each energy point in the Run III dataset. The relative values of the corrections applied to the \npot measurement as a function of the Run III period ID are shown in figure~\ref{fig:frontal_leakage}. Scan~1 and Scan~2 differ by approximately 3\%, because of the different vertical beam positions in the two scans.

\subsubsection{Radiation-induced losses}
\label{sec:ageing}

\begin{figure}[t]
    \centering
\includegraphics[width=0.49\linewidth]{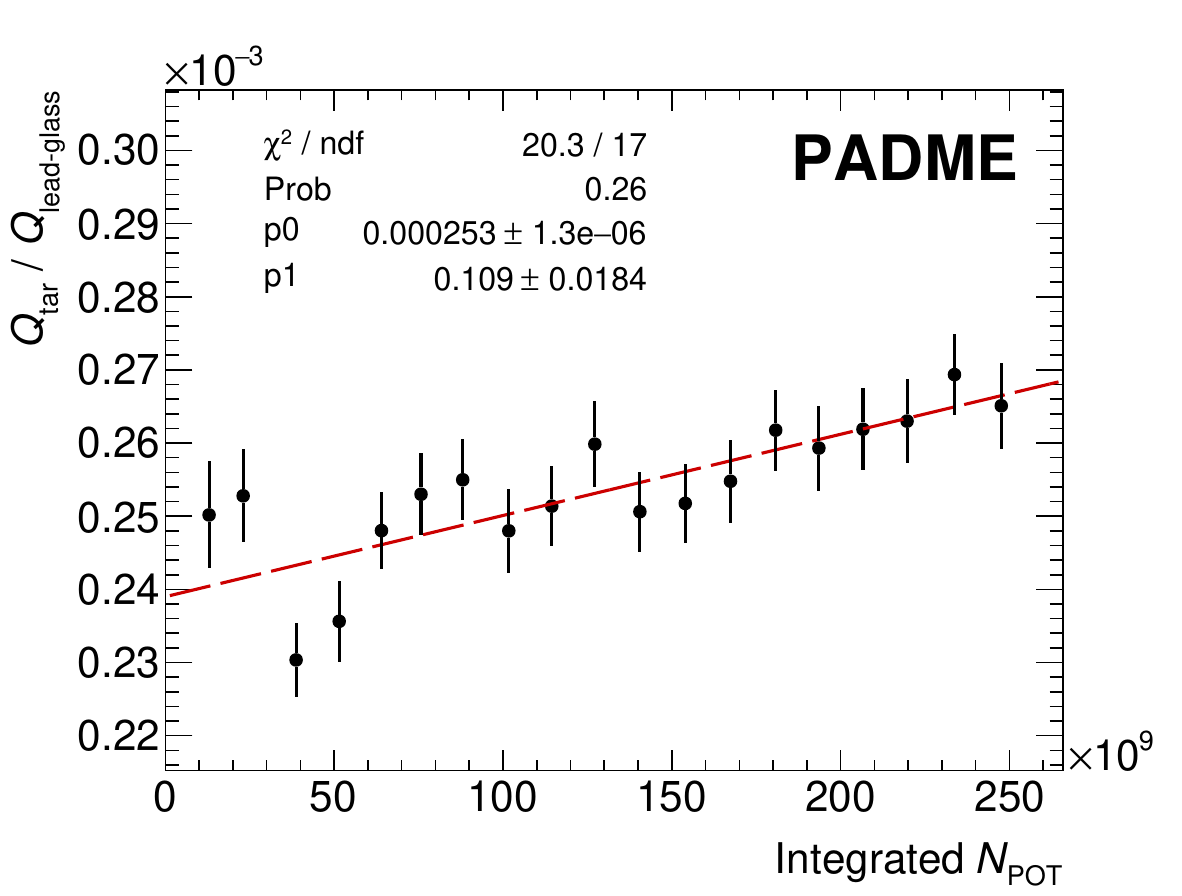}
    \includegraphics[width=0.49\linewidth]{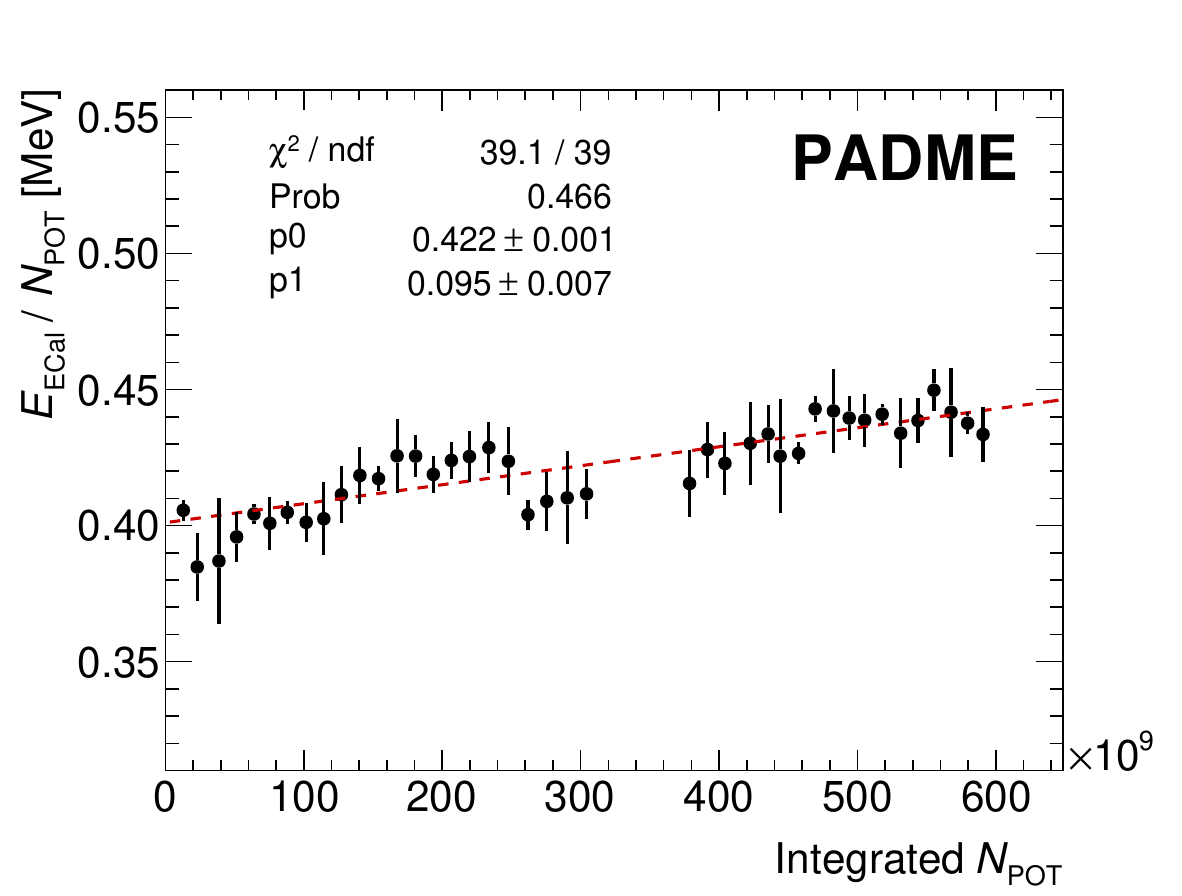}
        \caption{Ratio between the charge collected at the target \vs the lead-glass calorimeter (left), and between the average energy deposition measured by the ECal \vs the estimate of \npot (right), as functions of the integrated flux. Fits to the data according to eq.~(\ref{eq:fit-rad}) are overlaid.}
        \label{fig:Qx_Ehit}
\end{figure}

\begin{figure}[t]
    \centering        \includegraphics[width=0.6\textwidth]{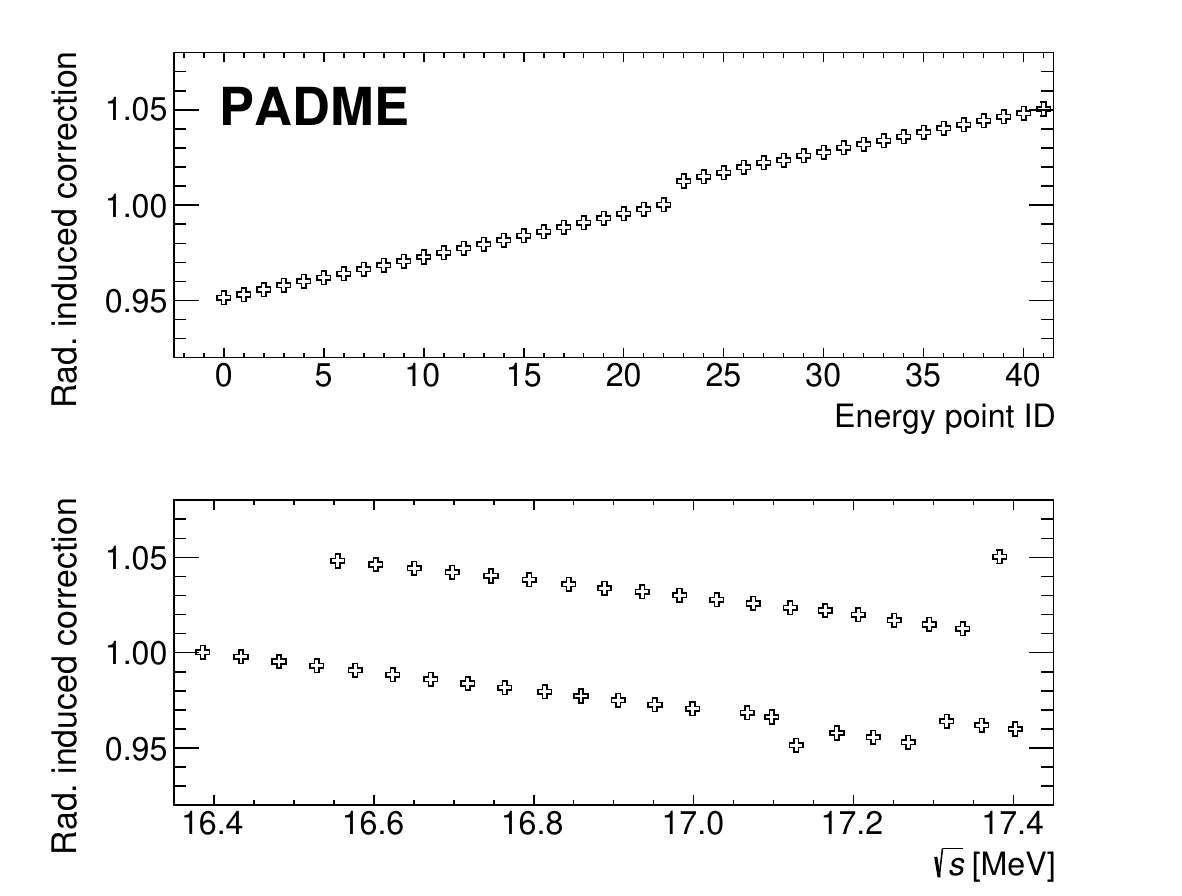}
        \caption{Relative values of the correction due to radiation-induced losses in the lead-glass calorimeter response as a function of the chronological ID of each energy point in the scan (top panel) and the \com energy (bottom panel).}
        \label{fig:ageingcorr}
    \hspace{0.05\textwidth}
\end{figure}

The transparency of the SF57 lead-glass calorimeter is affected by the accumulated dose~\cite{KOBAYASHI2001482-rad}. 
The total ionizing dose integrated during Run III in the first few radiation lengths of the detector is approximately 2.5\unit{krad}. A significant loss of transmittance is expected, especially for the short wavelengths where Cherenkov emission is dominant. Two proxy variables are used to derive the effect of this loss on the estimate of \npot: the charge $Q_\mathrm{tar}$ collected by the active target, and the mean value $E_\mathrm{ECal}$ of the total energy measured by the ECal. Both quantities are correlated with the flux, but each suffers from systematic uncertainties arising from the specific beam conditions. 
Nevertheless, both the ratio between $Q_\mathrm{tar}$ and the lead-glass charge (left panel of figure~\ref{fig:Qx_Ehit}) and that between $E_\mathrm{ECal}$ and \npot (right panel of figure~\ref{fig:Qx_Ehit}) provide a clear determination of the effect of the radiation-induced losses by showing a linear energy dependence on the integrated flux. The correction to be applied to \npot is determined by fitting each ratio $y$ with the function:

\begin{equation}
    y = p_0\times\left(1+p_1\times\frac{x-(x_F+x_I) /2}{x_F-x_I}\right),
    \label{eq:fit-rad}
\end{equation}

\noindent where $x_{I,F}$ are the  integrated fluxes for the first and last energy points of the scan. The $p_1$ results of the two fits in figure~\ref{fig:Qx_Ehit} are averaged, accounting for the relevant errors. Only Scan~1 estimates of $Q_\mathrm{tar}$ are used because of hardware issues that arose during Scan~2. The lead-glass calorimeter yield decreases by $0.097\pm0.007$ in Run III. This value is used to correct the estimated flux and its error is accounted for in the overall uncorrelated uncertainty budget. The relative uncertainty of the constant term ($p_0$) is 0.3\%, which contributes to the constant term uncertainty of \longK (also shown in table~\ref{tab:errorbudget}). The relative correction is shown in figure~\ref{fig:ageingcorr} as a function of the chronological ID of energy points and as a function of \com energy. 

\subsection{Determination of \texorpdfstring{\S}{S}}
\label{sec:signalYield}

\begin{figure}[t]
    \centering
    \includegraphics[width=0.55\linewidth]{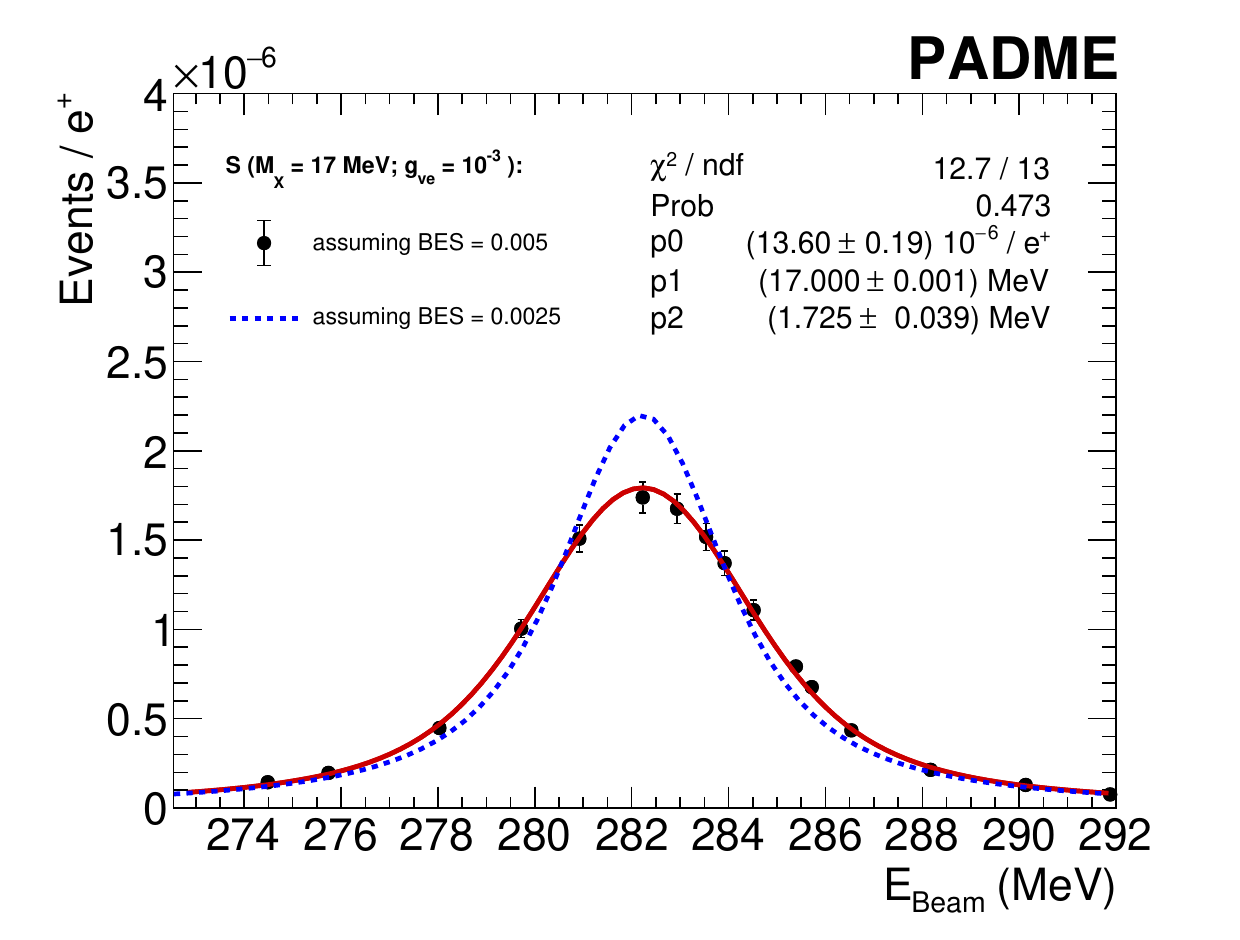}
    \caption{Signal yield \S as a function of beam energy, for $\MX = 17\MeV$ and $\gve=10^{-3}$. The open dots correspond to the assumption of a beam energy spread of 0.5\%, from~\cite{Arias-Aragon:2024qji}. A fit to a Voigt function is overlaid as a solid red line. The parameters of the Lorentzian component of the fit (peak height, center, and width) are listed in the inset. For comparison, the expected signal for a beam energy spread of 0.25\% is shown as the dotted blue line.}
    \label{fig:signal}
\end{figure}

The width of the expected event excess caused by the presence of the X17 particle, as a function of the \com energy, receives contributions from the spread of the beam energy and from the motion of the atomic electrons in the diamond target~\cite{Arias-Aragon:2024qji}. The signal shape published in ref.~\cite{Arias-Aragon:2024qji} for a relative beam energy spread of 0.5\% was parameterized as a Voigt distribution \vs beam energy. The parameters of the Lorentzian component of the distribution are listed in figure~\ref{fig:signal}. The beam-energy Gaussian spread was measured during data taking using the TimePix spot size in the horizontal coordinate, which corresponds to the bending plane used for momentum selection. The measured relative beam energy spread is $(0.25\pm0.05)\%$~\cite{Bertelli:2024ezd}. For comparison, the expected Voigt distribution with a beam energy spread of 0.25\% is also shown as the dashed line in figure~\ref{fig:signal}. Values and errors of the parameters of the Lorentzian distribution and of the beam energy spread are included in the probability density function of the related nuisance parameters in the statistical treatment discussed in section~\ref{sec:statTreat}.  

\begin{figure}[t]
    \centering
    \includegraphics[width=0.55\linewidth]{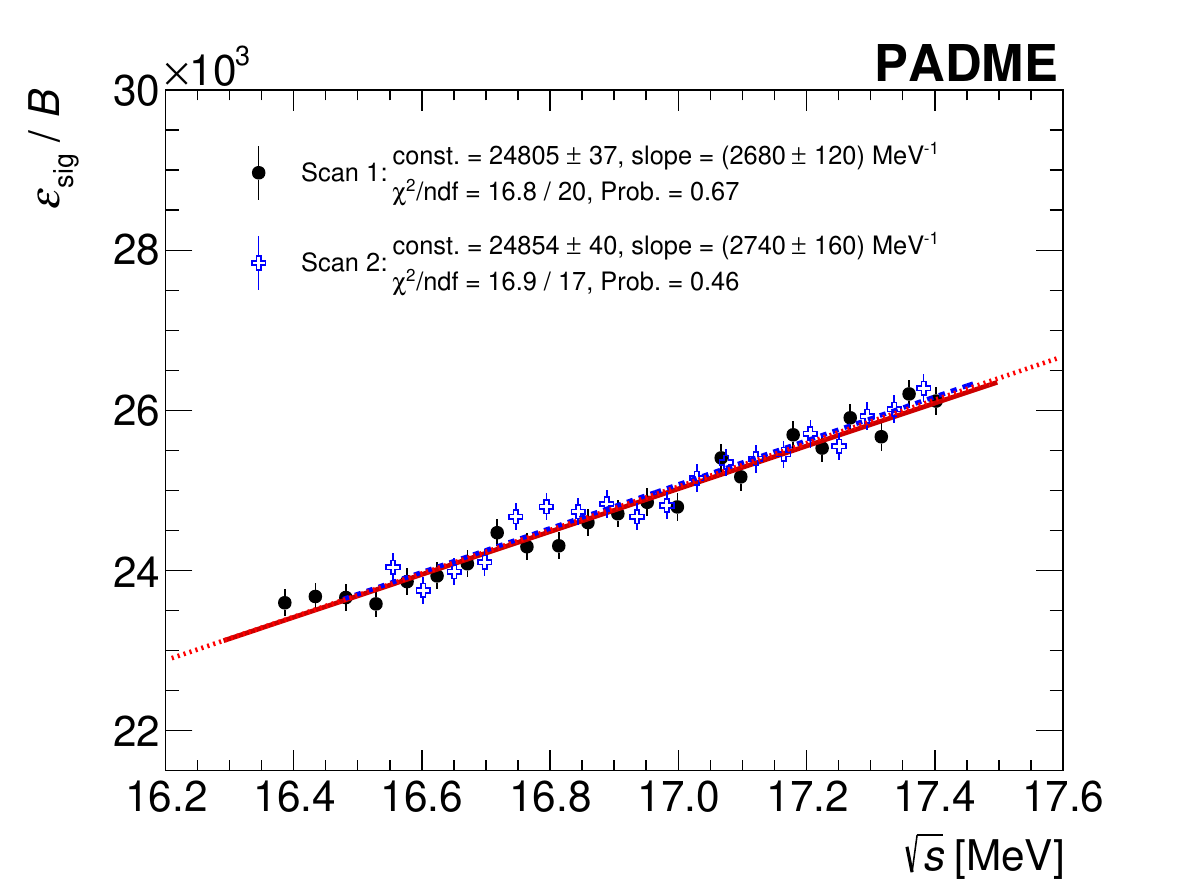}
    \caption{Ratio of signal efficiency \eps to SM background yield per \pot, \B, as a function of \com energy. Different symbols are used to represent the energy points of Scan~1 and Scan~2. Fits to linear functions performed separately for each scan are overlaid as the solid red and dashed blue lines, respectively. A combined fit using both scans is shown by the dotted red line.}
    \label{fig:epsoverB}
\end{figure}

The ratio of the signal efficiency, \longeps, to the expected SM background per \pot, \longB, contributes to the signal yield term in eq.~(\ref{eq:GRSignalSlope}). This ratio is shown in figure~\ref{fig:epsoverB} as a function of the \com energy, where only the statistical errors are included. The ratio $\eps/\B$ can be reliably parameterized with a separate linear function for each of the two scans. The parameters of the two linear functions are mutually compatible. Note that  fitting \eps ~and \B separately would provide incompatible parameters for the two scans. No additional source of systematic uncertainty is considered in the ratio $\eps/\B$, and a single linear parameterization is used for the statistical analysis, whose parameters are obtained from the combined fit of the two scans. The linear correlation between the constant term and the energy slope is about $-1.8\%$.

\section{The unblinding procedure}
\label{sec:blindunblinding}

Given the expected uncertainty in the X17 mass and the signal width induced by the electron motion and by the beam-energy spread, no sideband region in the \sqrts distribution can be  defined in advance. Nevertheless, a validation procedure~\cite{Bertelli:2025mil} was designed to perform blind data analysis while still allowing control of residual systematic uncertainties and data consistency.

Since values of the X17 coupling strength \gve above $10^{-3}$ have already been excluded~\cite{Anastasi:2015qla},
a sideband region can be defined by optimizing the quality of a linear fit to \longgR of eq.~(\ref{eq:GRBkgSlope}), performed excluding a fixed number (10) of scan energy points. The center of the excluded region is determined by an automatic optimization procedure and is not revealed to the analyzers. The data validation criteria are based on the fit quality, the distribution of the fit pulls, and the parameters of the linear fit. 

The following criteria, with the data still blinded, must be sequentially satisfied  before unblinding can proceed. If any of the checks fail, the unblinding must be aborted.

\begin{enumerate}
    \item Fit quality: The optimized linear fit
    of \longgR must have a minimum $\chi^2$ corresponding to a probability (cumulative integral of the $\chi^2$ distribution) P$(\chi^{2}>\chi^2_\mathrm{min}\mbox{; ndf}) > 20\%$, where ndf is the number of degrees of freedom.
    \item Fit parameter reliability:  The fit parameters (constant and energy-dependent slope of \longK) are required to be within two standard deviations (a-priori systematic uncertainty) of their expected values.
    \item Gaussian fit pulls: The distribution of the fit pulls is unmasked. It must be consistent with a Gaussian distribution, namely the probability P$(\chi^2\mbox{; ndf})$ for a Gaussian fit to the pulls distribution must be higher than 5\%. 
    \item Correlated systematic errors on the fit pulls: The data in the sideband region is unmasked. The sequence of pulls in the sidebands is joined in a single continuous 
    sequence, such that the location of the masked region still cannot be uncovered. The points are sorted according to the data taking acquisition time and the possible gaps are of unknown extent and at unknown locations. The sequence is then fit to a straight line and the resulting slope must be consistent with zero within two standard deviations.     
\end{enumerate}

If any of the above criteria were not satisfied, the analysis technique would have needed revision. Details on the above criteria are discussed in the following subsections.

\begin{figure}[t]
    \centering
    \includegraphics[width=0.49\linewidth]{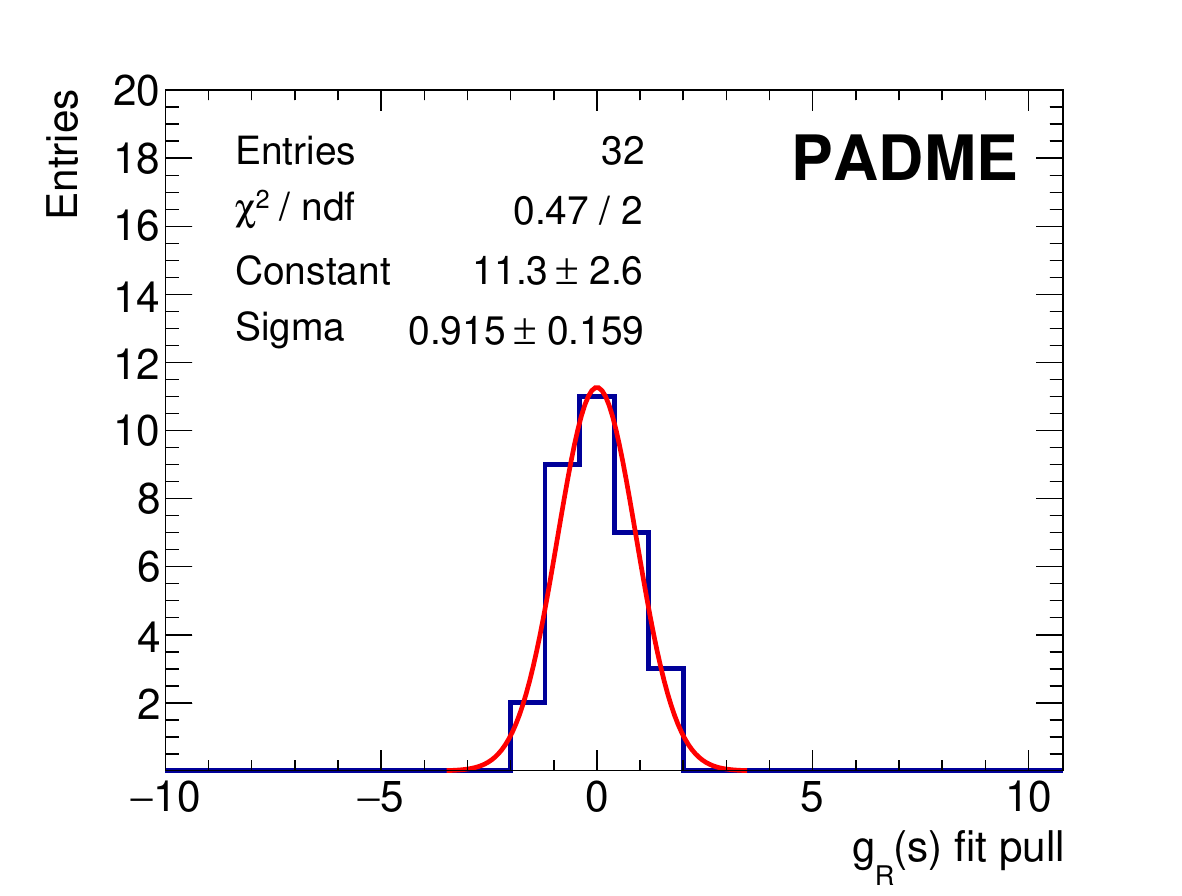}   
\includegraphics[width=0.49\linewidth]{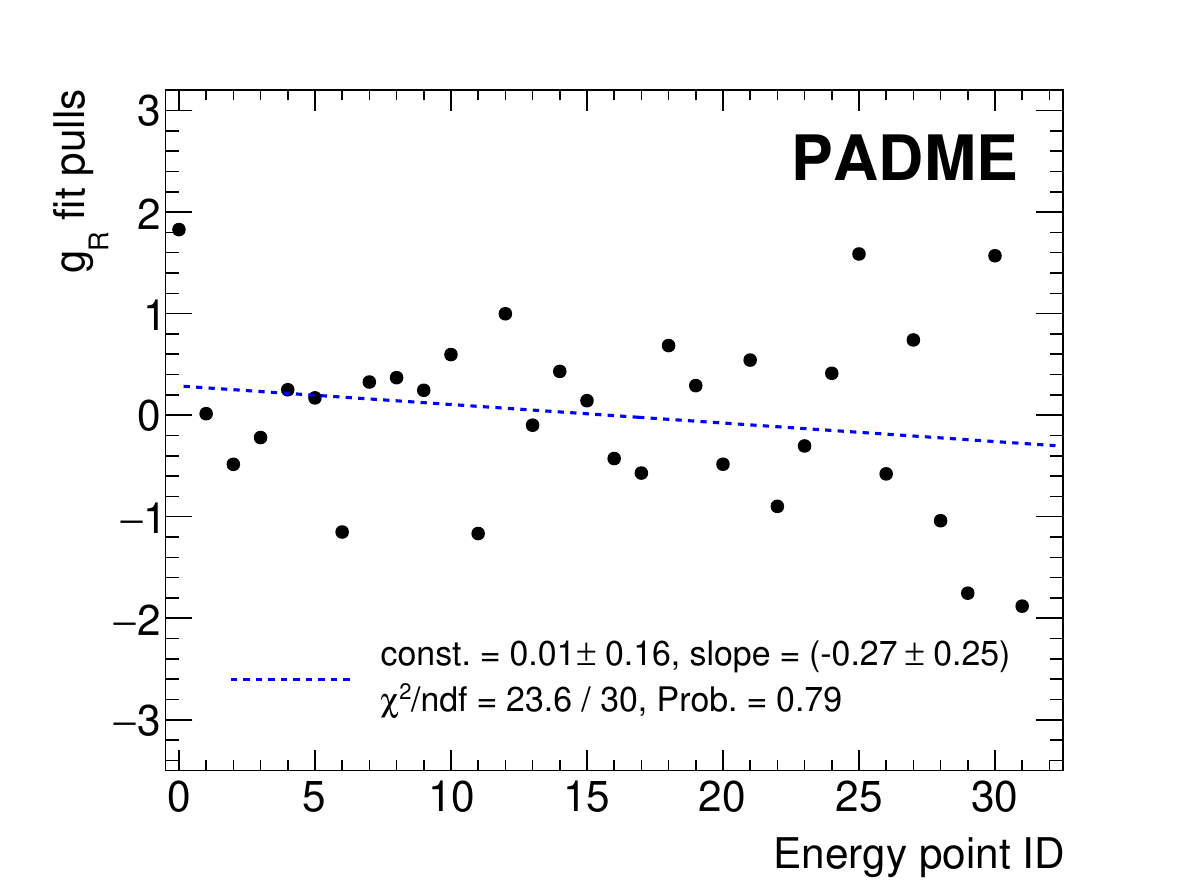}
    \caption{Pull distributions for a linear fit to \longgR after masking a region of 10 scan points (left). Fit pulls \vs chronological ID of the energy points (right). A Gaussian (linear) fit is superimposed on the left (right) panel.}
    \label{fig:pulls}
\end{figure}

The procedure described was applied to the blinded data, leading to the  following results:

\begin{enumerate}
    \item Fit quality: P$(\chi^2>\chi^2_\mathrm{min}) = 0.74$.
    \item Fit parameter reliability: The constant term and slope parameters are $1.012\pm0.002$ and $(-0.010\pm0.005)\MeV^{-1}$. The results are compatible within two standard deviations with the expected values $1.00\pm0.028$ and $(-0.006\pm0.006)\MeV^{-1}$, respectively.
    \item Gaussian fit pulls: The fit pulls are shown in figure~\ref{fig:pulls}. The Gaussian fit, with a mean fixed to 0, has a probability of 80\%.
    \item Correlated systematic on the fit pulls: The pull-sequence fit determines a slope parameter $1.08\sigma$ away from zero.
\end{enumerate}

\subsection{Fit quality condition}
The cut value in condition 1 is a  compromise: on the one hand, from S+B simulations, we determined that the threshold applied in condition 1 corresponds to rejecting about 10\% of the experiment outcomes for a signal coupling $\gve = 10^{-3}$. On the other hand, for B-only scenarios, the distribution of $\chi^2_\mathrm{min}$ is found to be biased towards small values by the removal of the masked region, so that it does not follow a proper $\chi^2$ distribution for ndf degrees of freedom. In absence of signal, the probability to randomly reject an experiment outcome because of condition 1 is indeed below 2\%. 

\begin{figure}[t]
    \centering
    \includegraphics[width=0.55\textwidth]{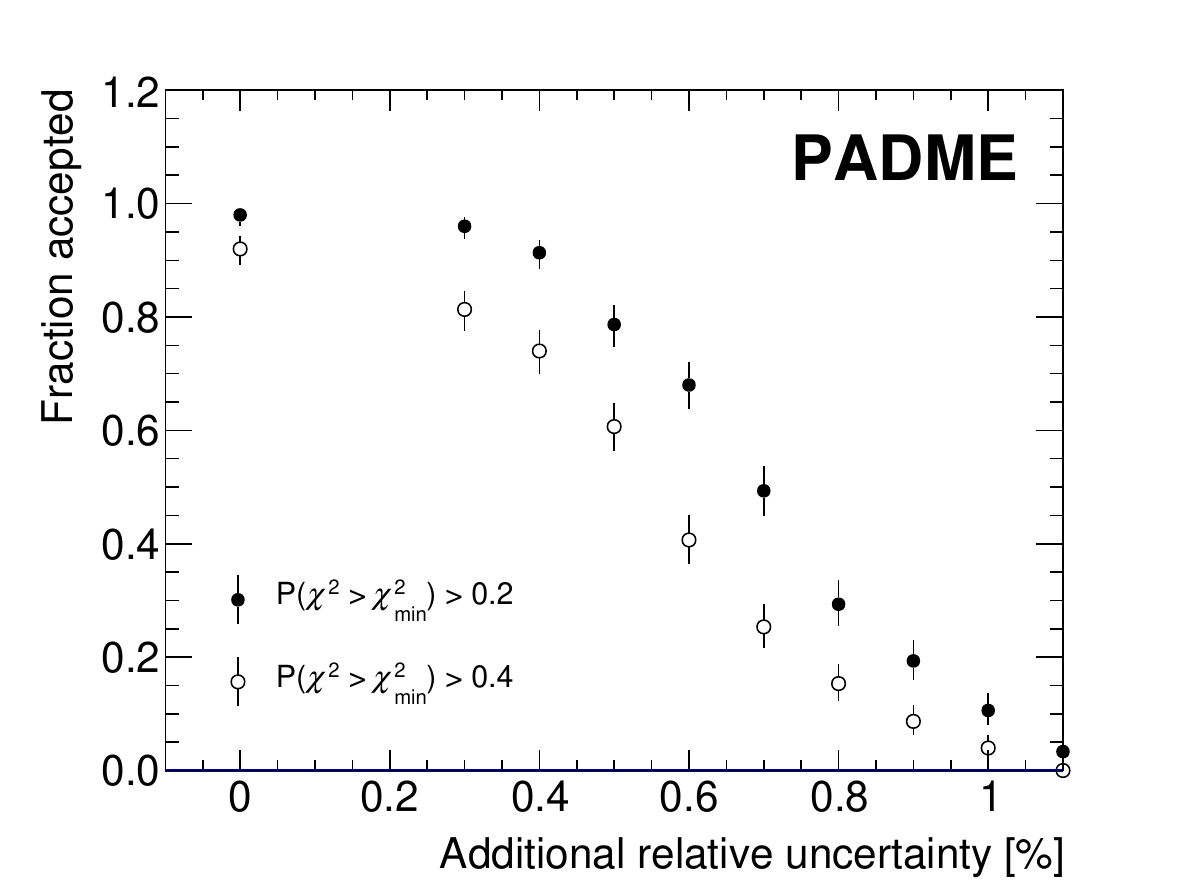}
        \caption{Effect of an additional relative uncertainty on the fraction of accepted experimental outcomes as seen with B-only simulations of \longgR. The additional relative uncertainty is added in quadrature to the nominal uncertainty from the analysis (approximately 0.9\%), which increases the statistical fluctuations. Solid points correspond to the nominal probability criterion applied in the unblinding, and open points to a stricter requirement.}
        \label{fig:passprobe}
\end{figure}

Fulfilling the fit quality condition allows the estimation of additional uncorrelated systematic uncertainties beyond those identified in the main analysis, reaching approximately 0.9\% per energy point. To determine the constraining power of the fit quality condition, B-only simulations
were performed where the total fluctuations of \longgR include a further relative uncertainty added in quadrature to the estimated value. As the true fluctuations depart further from the estimated value, the fit-quality condition 1 retains fewer experimental outcomes, as shown in figure~\ref{fig:passprobe}. Given the condition applied (solid points), fulfilling condition 1 implies that, at 90\% confidence level (\CL), no additional uncorrelated fluctuations stronger than 1\% can be present in the data. Applying a tighter condition such as $P(\chi^2>\chi^2_\mathrm{min})>0.4$ would have constrained additional unknown sources of uncorrelated fluctuation to below 0.8\% at 90\% \CL, but at the cost of allowing approximately 10\% of the experiment outcomes without additional uncertainties to be rejected solely due to statistical fluctuations. Moreover, the presence of signal with  $\gve\simeq8\times10^{-4}$ (as opposed to $\gve\simeq 10^{-3}$ for the condition used) would have generated additional random rejections with 5--10\% probability. Thus the cut chosen for condition 1 preserves the concept of a data sideband in the presence of a signal with coupling strength up to $\gve\simeq 10^{-3}$, while failing to be fulfilled in only a few percent of the experiments, for B-only scenarios. Satisfying condition 1 implies an absence of additional sources of \sqrts-uncorrelated relative uncertainties above 1\% at 90\% \CL.

\subsection{Fit parameter reliability}
\begin{figure}[t]
    \centering
    \includegraphics[width=0.55\textwidth]{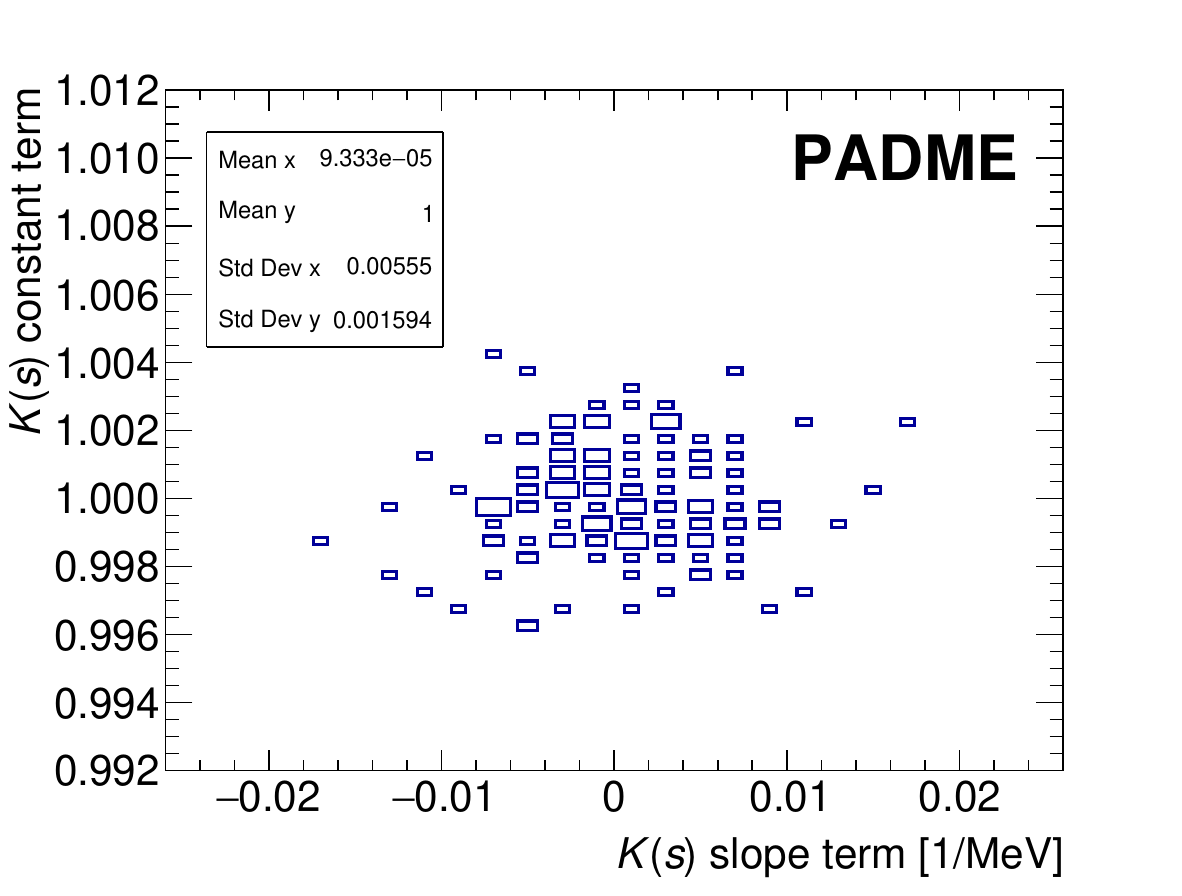}
        \caption{Distribution of linear fit parameters obtained by simulating the B-only scenario of \longgR. The $y$-axis shows the constant term of \longK, and the $x$-axis the slope term.}
        \label{fig:fitparameters}
\end{figure}

The analysis provides a-priori estimates for the parameter values (constant term and \com-energy slope of \longK) and their associated uncertainties. Condition 2 requires consistency of the fit results with the a-priori estimates, within $2\sigma$ intervals. The fit parameters obtained for simulated B-only scenarios with $\longK = 1$ are shown in figure~\ref{fig:fitparameters}. The fit results are found to provide unbiased estimates of the true values: 1~and~ 0~MeV$^{-1}$, respectively, for the constant and slope terms in this scenario. The correlation coefficient of the fit parameters is  $-0.1$ or less. The expected standard deviations of constant and slope terms are less than 0.002 and around $0.005\MeV^{-1}$, respectively. Therefore, the fit results $1.012\pm 0.002$ and
$(-0.010\pm0.005)\MeV^{-1}$ are statistically significant. 

\section{Statistical treatment of the data}

\label{sec:statTreat}
A detailed description of the statistical technique used for the determination of the expected sensitivity is given in ref.~\cite{Bertelli:2025mil}.
For a given mass hypothesis \MX, the procedure yields upper limits on the coupling strength \gve by scanning a predefined set of test \gve values. A modified frequentist approach known as \CLs is used~\cite{ATLAS:2011tau}. The test statistic compares the likelihood functions between the S+B and the B-only scenarios~\cite{Junk:2005awa,Junk:2006fye}. These functions depend on a set of nuisance parameters $\theta$.

The likelihood function is defined as:

\begin{equation}
    L(\text{data} | \MX, \gve, \theta) =  \left(\prod_{s_i} \frac{1}{\sqrt{2\pi \sigma_{g_R(s_i)}^2}} e^{-\frac{(g_R(s_i) - g_{R,\mathrm{exp}}(s_i))^2}{2\sigma_{g_R(s_i)}^2 }}\right) \times P(\theta),
\end{equation}

\noindent where \gR is the main analysis observable as defined in eq.~(\ref{eq:GR}) and the expected values $g_{R,\text{exp}}$ are defined in eqs.~(\ref{eq:GRBkgSlope})~and~(\ref{eq:GRSignalSlope}) for the B-only and S+B scenarios, respectively, and $P(\theta)$ is the probability for a particular set of nuisance parameters. In B-only scenarios, two nuisance parameters are used: the constant and the slope terms of \longK. In S+B scenarios, five additional nuisance parameters enter the calculation: three related to the signal shape (the normalization of the signal yield, the width of the Lorentzian component of the signal model, and the beam energy spread), and two related to the ratio between the signal efficiency and the background yield per \pot (the constant and the slope terms of a linear parametrization \vs \sqrts). Initial expected values and uncertainties for each nuisance parameter enter the $P(\theta)$ calculation. The probability density functions for the nuisance parameters are taken to be either a single Gaussian function or a multi-dimensional Gaussian function if correlation terms must be accounted for (which is the case for the \longK and $\longeps/\longB$ parameterizations).

    \label{eq:statistic}

The value of $\CLs(\MX,\gve)$ is computed as the ratio between the probabilities for the S+B ($p_s$) and B-only ($p_b$) hypotheses, obtained via pseudo-experiments:

\begin{equation}
    \CLs(\MX,\gve) = \frac{p_s(\MX,g_{ve})}{1 - p_b}.
\end{equation}

In the pseudo-experiment procedure, the observables entering \longgR (\Ntwo, \npot, and \B) are sampled around their observed values while the nuisance parameters are frozen to the values $\hat{\theta}$ obtained from the likelihood minimization~\cite{ATLAS:2011tau}.
If $\CLs(\MX,\gve) < \alpha$, then, 
for a given value of \MX, values of the coupling strength higher than \gve are excluded with $(1-\alpha)$ confidence level.

\subsection{Expected sensitivity}

To estimate the expected sensitivity, more than 300 experimental outcomes were generated. For each outcome, both observables and nuisance parameters were sampled around their expected central values while accounting for their uncertainties.

\begin{figure}[t]
    \centering \includegraphics[width=0.49\textwidth]{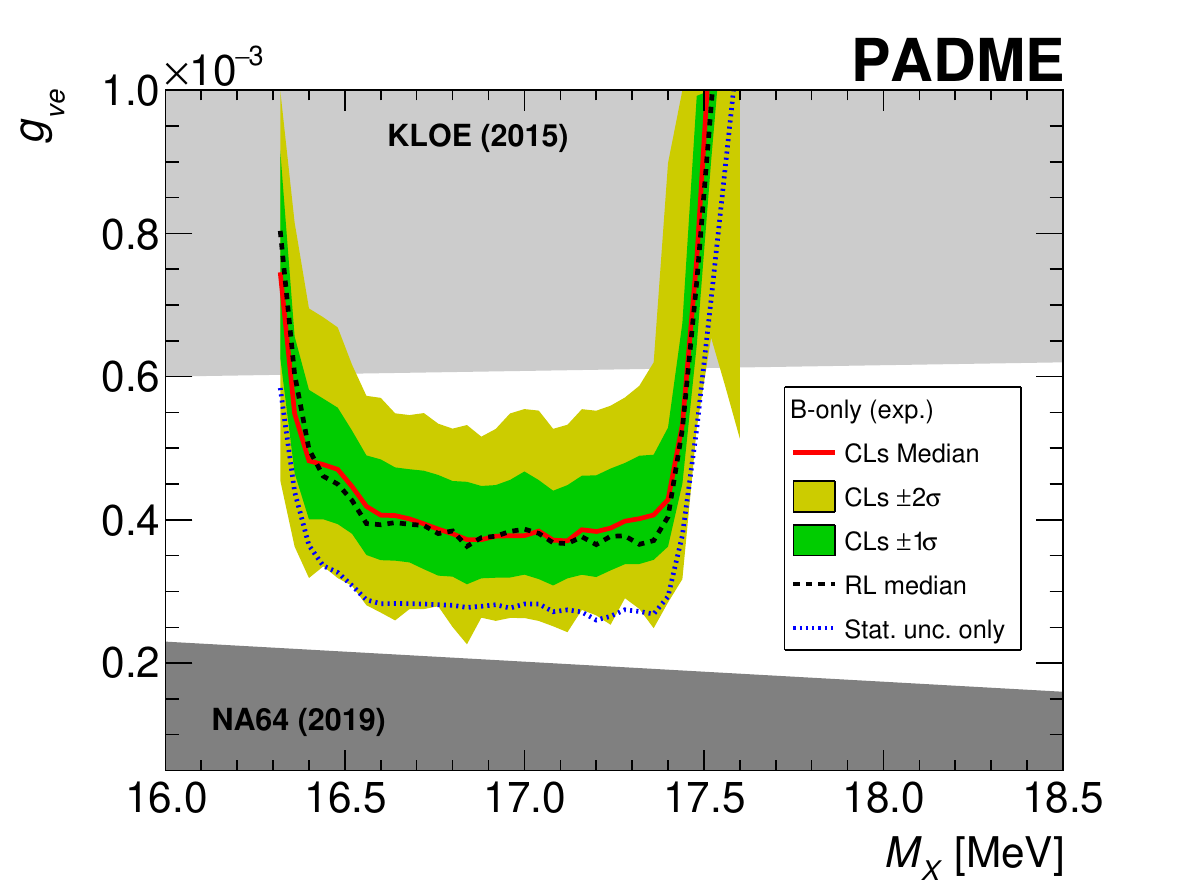}
\includegraphics[width=0.49\textwidth]{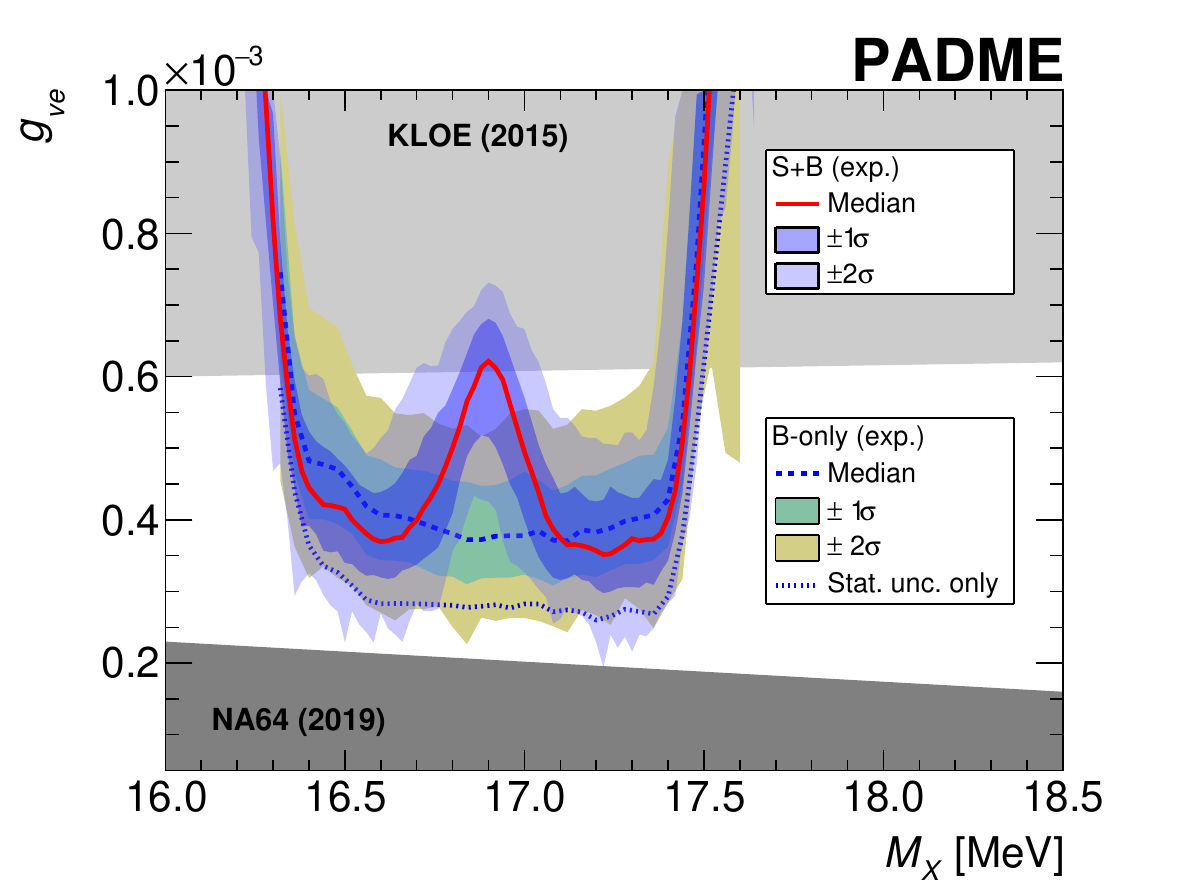}
       \caption{Expected exclusion upper limits at 90\% \CL in the absence of a signal (left), and in the presence of a signal with $\gve=5\times 10^{-4}$ and $\MX=16.9\MeV$ (right), in addition to the predicted background. The left panel also shows the median expected limit computed with the Rolke log-likelihood (RL) method~\cite{Rolke:2000ij,Rolke:2004mj}.}
       \label{fig:MCsens} 
\end{figure}

The left panel of figure~\ref{fig:MCsens} shows the expected exclusion limit at 90\% \CL in the absence of signal. The red line indicates the median upper limit, and green and yellow bands represent the ${\pm}1\sigma$ and ${\pm}2\sigma$ coverages. The median closely matches the result from the Rolke log-likelihood method~\cite{Rolke:2000ij,Rolke:2004mj}, shown as a dashed line (RL) in the figure. Including all systematic uncertainties leads to weaker expected limits than those arising purely from B-only fluctuations, denoted by the dotted blue line. Compared to the sensitivity reported in ref.~\cite{Bertelli:2025mil}, the present result shows an improvement that is primarily due to reduced uncorrelated uncertainties on \longgR.

The right panel of figure~\ref{fig:MCsens} shows the median upper limit in the presence of a signal assuming $\gve = 5\times10^{-4}$ and $\MX = 16.9\MeV$ (red line). The $1\sigma$ and $2\sigma$ exclusion bands for such a S+B scenario are overlaid as blue filled regions onto the B-only exclusion bounds (green and yellow regions) and the B-only median upper limit (dashed blue line).

\section{Results and post-unblinding checks}


    


\begin{figure}[t]
    \centering
       \centering        \includegraphics[width=0.55\textwidth]{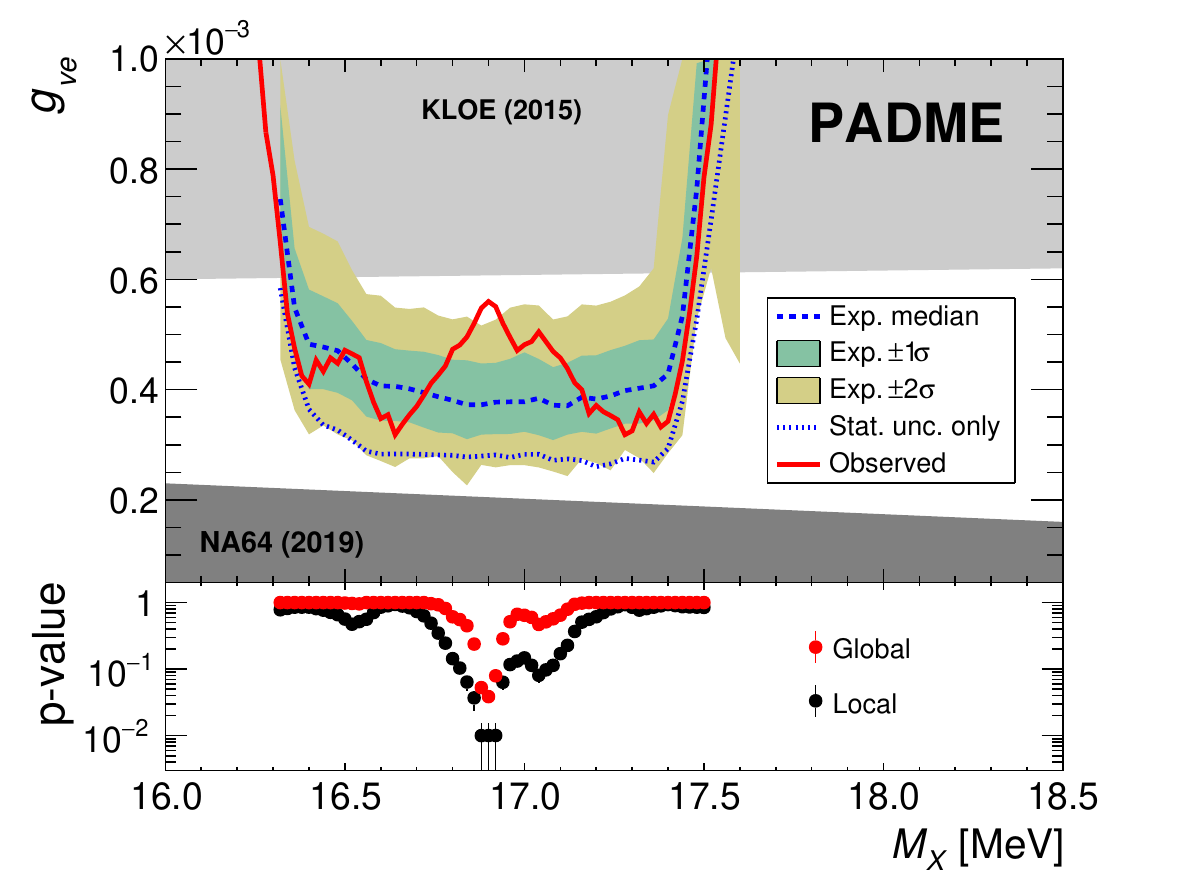}
        \caption{Exclusion upper limits as a function of \MX at 90\% \CL. The solid red line is the observed limit curve, and the green and yellow bands are the $1\sigma$ and $2\sigma$ local envelopes around the median expected curve (dashed blue line) for the B-only hypothesis. The gray filled areas are the excluded regions at 90\% \CL from the KLOE~\cite{Anastasi:2015qla} and NA64~\cite{NA642019} experiments. The bottom panel shows the probability values corresponding to the observed upper limits. Black and red dots denote local and global probabilities, respectively.}
        \label{fig:BoxOpen}
\end{figure}

After the unblinding procedure, the \CLs method was applied to data to determine the final observed upper limit. The results are shown in figure~\ref{fig:BoxOpen}. 
The observed upper limit at 90\% \CL (solid red line) is weaker than the B-only expectation, exceeding the $2\sigma$ local coverage for $\MX =16.90\MeV$ and $\gve = 5.6\times 10^{-4}$. The local probability (black dots in the bottom panel) shows a corresponding minimum of 1\% (a $2.5\sigma$ effect).

The global probability to obtain such an excess assuming the B-only hypothesis was evaluated via simulation. The range between the median value of the coupling strength and the $2\sigma$ upper band (multiplied by 3) was split into 48 ``test points''. The fraction of times a B-only upper limit exceeds a test point anywhere in the mass interval of interest ($16.6<\MX<17.2\MeV$) is an estimator of the global probability. The global probability for the observed limit (red points in the bottom panel of figure~\ref{fig:BoxOpen}) is calculated by interpolating the results from the test points. The minimum global probability is $3.9_{-1.1}^{+1.5}\%$, corresponding to an excess of $(1.77 \pm 0.15)\sigma$. A reduced mass interval of $16.85\pm0.12\MeV$, corresponding to a ${\pm}3\sigma$ range around the combined fitted mass as reported in ref.~\cite{Denton:2023gat}, would lead instead to a global p-value of $(2.00\pm0.19)\sigma$. 

\begin{figure}[t]
\centering\includegraphics[width=0.55\linewidth]{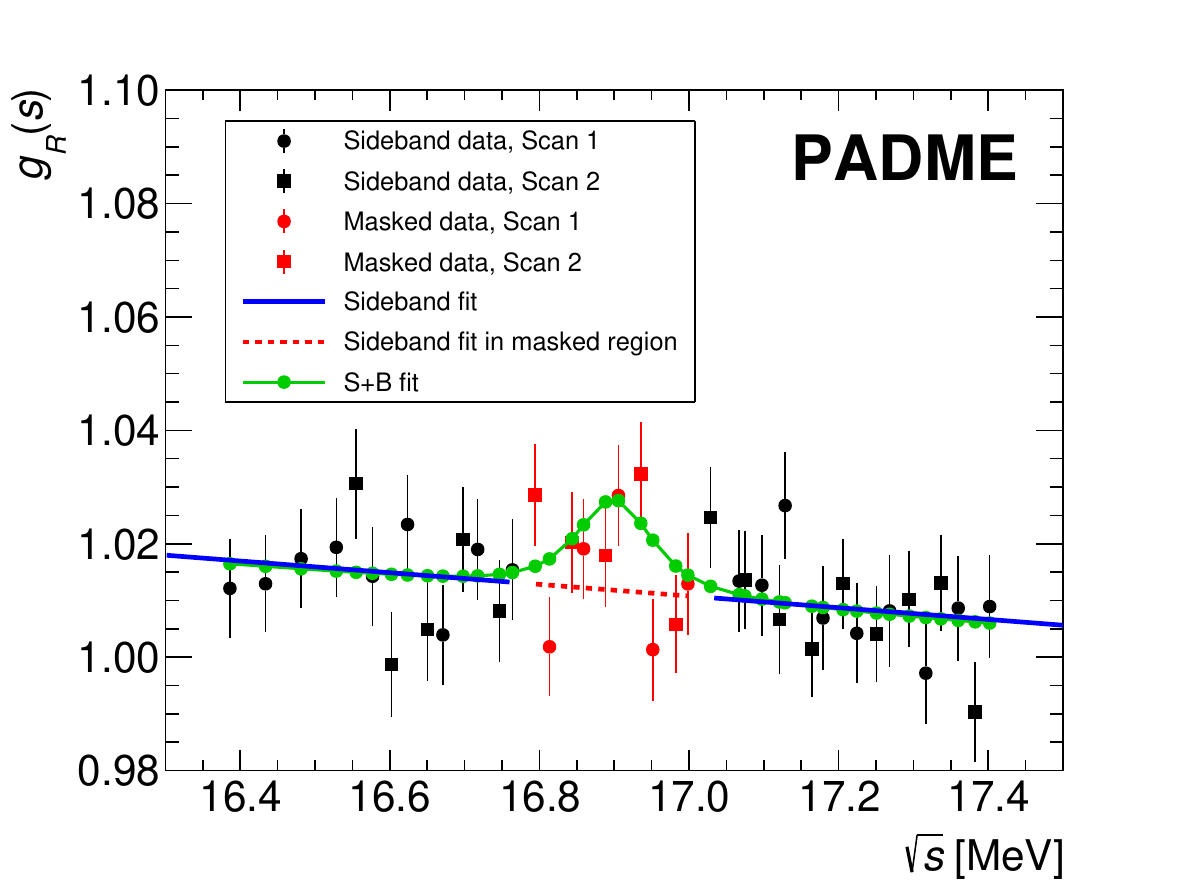}
    \caption{Observed distribution of \gR as defined in eq.~(\ref{eq:GR}) \vs \sqrts. Circle (square) points denote the data collected in Scan~1 (Scan~2). Red points denote the region masked by the automatic unblinding procedure. The solid blue curve represents a fit to the sideband points, assuming only background contributions. The dashed red line interpolates the sideband fit into the masked region. Finally, the solid green line denotes the full S+B fit to the data. }
    \label{fig:gRobs}
\end{figure}

For illustration purposes, the distribution of \longgR is shown as a function of \sqrts in figure~\ref{fig:gRobs}. The data from Scans~1~and~2 are represented by different markers, with the energy points that were masked by the ``blind unblinding'' procedure described in section~\ref{sec:blindunblinding} shown in red. The S+B fit to the observed data corresponding to the minimum global p-value is shown with green points.

A second excess is present at $\MX\simeq17.1\MeV$, corresponding to a local (global) probability of approximately 8\% (40\%).

\subsection{Post-unblinding checks}
After data unblinding, a number of systematic checks were also carried out:

\begin{itemize}
    \item An alternative evaluation of the statistical significance was performed using the \CLsb statistic and the power-constrained limit method. The result shows a weaker upper limit in the same mass point as in the nominal \CLs-based analysis, with only a slightly weaker global significance of $(1.63\pm0.13)\sigma$;
    \item An alternative form for the \longK correction in eqs.~(\ref{eq:GRBkgSlope})~and~(\ref{eq:GRSignalSlope}) was assumed, corresponding to the absence of a \com energy slope. The result again shows a weaker upper limit at the same mass value observed in the nominal analysis. The statistical global significance slightly lowers to $(1.60\pm0.15)\sigma$; 
    \item An alternative correction for radiation-induced losses was implemented. The uncertainty related to the slope as a function of the integrated flux was treated as a separate nuisance parameter instead of an uncorrelated error per energy point. The statistical global significance was only slightly reduced to $(1.65\pm0.13)\sigma$;
    \item The correction due to radiation-induced losses can be verified ex-post by fitting the uncorrected \gR values as a function of the integrated positron flux. Using a linear fit similar to eq.~(\ref{eq:fit-rad}), a slope of $0.10\pm0.04$ was computed, in excellent agreement with the determination in section~\ref{sec:ageing}. 
\end{itemize}

\section{Conclusions}

We presented the results of a search for the hypothetical X17 particle with the PADME Experiment. The analysis relies on the measurement of the cross section for the production of events with a two-body final state (either $e^+e^-$ or $\gamma\gamma$) from $e^+e^-$ annihilation processes with a center-of-mass energy range of 16.4--17.4\MeV. A blind analysis is performed, and the total uncorrelated uncertainty for each of the 42 energy points is shown to be below 1\%. Upper limits on the coupling strength of the hypothetical particle are set in previously unexplored regions of the available parameter space.
An excess of events over the predicted background expectation is observed for an X17 mass of $16.90\MeV$, with local and global significance of 2.5 and $1.8\pm0.2$ standard deviations, respectively. 
The location of this excess is consistent with the average value of the results reported by the ATOMKI experiments.

The PADME Collaboration has started a new data-taking campaign in 2025 with an upgraded detector, aiming to improve the experimental sensitivity for the same mass window presented in this paper. Data points corresponding to several energy values have already been collected, and data taking is foreseen to continue until the end of 2025.

\section*{Acknowledgments}

The PADME Collaboration acknowledges significant support from the Istituto Nazionale di Fisica Nucleare, and in particular the Accelerator Division and the LINAC and BTF teams of INFN Laboratori Nazionali di Frascati, for providing an excellent quality beam and full support during the data collection period. 
The authors are also grateful to E. Nardi, L. Darmé, G. Grilli di Cortona, and F. Arias-Aragón for clarifying the role of atomic electron motion on the signal production yield~\cite{Arias-Aragon:2024qji}.
The authors also thank 
C. Carloni Calame for his advice on estimating the NLO Bhabha cross section and kinematics using the Babayaga generator.
Sofia University team acknowledges 
support
by the European Union - NextGenerationEU, 
through the National Recovery and Resilience Plan of the Republic of Bulgaria
project SUMMIT BG-RRP-2.004-0008-C01, 
by BNSF KP-06-COST/25 from 16.12.2024 based upon
work from COST Action COSMIC WISPers CA21106 
supported by COST (European Cooperation in Science and Technology), 
and 
by TA-LNF as part of STRONG-2020 EU Grant Agreement 824093.

\bibliographystyle{JHEP}
\bibliography{bibliography}
\end{document}